\documentclass[ALICE,manyauthors]{cernphprep}

\newcommand{\kstar}{$k^{*}$\xspace}
\newcommand{\ktrue}{$k^{*}_{\mathrm{True}}$\xspace}
\newcommand{\krec}{$k^{*}_{\mathrm{Rec}}$\xspace}
\newcommand{\minv}{$m_{\mathrm{inv}}$\xspace}
\newcommand{\mt}{$m_{\mathrm{T}}$\xspace}
\newcommand{\pt}{$p_{\mathrm{T}}$\xspace}

\newcommand{\Lam}{$\Lambda$\xspace}
\newcommand{\ALam}{$\overline{\Lambda}$\xspace}
\newcommand{\LamALam}{$\Lambda$ ($\overline{\Lambda}$)\xspace}

\newcommand{\KchP}{$\mathrm{K^{+}}$\xspace}
\newcommand{\KchM}{$\mathrm{K^{-}}$\xspace}
\newcommand{\Kpm}{$\mathrm{K^{\pm}}$\xspace}

\newcommand{\Ks}{$\mathrm{K^{0}_{S}}$\xspace}

\newcommand{\LamK}{$\Lambda$K\xspace}

\newcommand{\LamKchP}{$\Lambda\mathrm{K^{+}}$\xspace}
\newcommand{\ALamKchM}{$\overline{\Lambda}\mathrm{K^{-}}$\xspace}

\newcommand{\LamKchM}{$\Lambda\mathrm{K^{-}}$\xspace}
\newcommand{\ALamKchP}{$\overline{\Lambda}\mathrm{K^{+}}$\xspace}

\newcommand{\LamKpm}{$\Lambda\mathrm{K^{\pm}}$\xspace}

\newcommand{\LamKs}{$\Lambda\mathrm{K^{0}_{S}}$\xspace}
\newcommand{\ALamKs}{$\overline{\Lambda}\mathrm{K^{0}_{S}}$\xspace}

\newcommand{\XiKchP}{$\Xi^{-}\mathrm{K^{+}}$\xspace}

\newcommand{\XiKpm}{$\Xi^{-}\mathrm{K^{\pm}}$\xspace}

\newcommand{\Vz}{V$^{0}$\xspace}
\usepackage[comma,square,numbers,sort&compress]{natbib}
\usepackage{color}	%May be necessary for colored links
\usepackage{hyperref}
\usepackage{lineno}

\usepackage[T1]{fontenc}

\usepackage{multirow}
\usepackage{boldline}  % to make lines in table bold

\usepackage{pdflscape}

\usepackage{comment}

\usepackage{subfiles}
\usepackage{chngpage}  % for adjustwidth
\usepackage{boldline}  % to make lines in table bold
\usepackage{arrayjobx} % To use the array structures stored in FitResults_cLamcKch_20180505.tex 

\usepackage{relsize} %for \mathlarger
\usepackage{scalerel}  %to use scaleto functionality for math eqns (ex. in exponent of lambda eqn)

\usepackage{upgreek} %for \uppi, etc.

\hypersetup{
	linktoc=all,
	colorlinks=false,
	linkbordercolor={1 0 0},
	citebordercolor={0 1 0},
	filebordercolor={0 .5 .5},
	menubordercolor={1 0 0},
	runbordercolor={0 .7 .7},	
	urlbordercolor={0 1 1}
}

\begin{document}%

%*************************************************  Title page **********************************************************
%************************************************************************************************************************
\begin{titlepage}
\PHyear{2020} % CERN-EP-2020-080
\PHnumber{080}      % required, will be obtained from PH
\PHdate{19 May}  % required, will be obtained from PH
%

%%% Put your own title + short title here:
\title{\LamK femtoscopy in Pb--Pb collisions at $\mathbf{\sqrt{{\textit s}_{NN}}} =$ 2.76 TeV}
\ShortTitle{\LamK femtoscopy in Pb--Pb collisions}   % appears on right page headers

%%% Do not change the next lines
\Collaboration{ALICE Collaboration\thanks{See Appendix~\ref{app:collab} for the list of collaboration members}}
\ShortAuthor{ALICE Collaboration} % appears on left page headers, do not change

\begin{abstract}
%%%%%%%%%%%%%%%%%%%%%% Do not use any of the shorthand definitions in the abstract %%%%%%%%%%%%%%%%%%%%%%%%%%%%%%%%%%%
The first measurements of the scattering parameters of $\Lambda$K pairs in all three charge combinations ($\Lambda$K$^{+}$, $\Lambda$K$^{-}$, and $\Lambda\mathrm{K^{0}_{S}}$) are presented.
The results are achieved through a femtoscopic analysis of $\Lambda$K correlations in Pb--Pb collisions at $\sqrt{s_{\mathrm{NN}}}$ = 2.76 TeV recorded by ALICE at the LHC.  
The femtoscopic correlations result from strong final-state interactions, and are fit with a parametrization allowing for both the characterization of the pair emission source and the measurement of the scattering parameters for the particle pairs.
Extensive studies with the THERMINATOR 2 event generator provide a good description of the non-femtoscopic background, which results mainly from collective effects, with unprecedented precision.
Furthermore, together with HIJING simulations, this model is used to account for contributions from residual correlations induced by feed-down from particle decays.
The extracted scattering parameters indicate that the strong force is repulsive in the \LamKchP interaction and attractive in the \LamKchM interaction.
The data hint that the \LamKs interaction is attractive, however the uncertainty of the result does not permit such a decisive conclusion.
The results suggest an effect arising either from different quark--antiquark interactions between the pairs ($\rm s\overline{s}$ in $\Lambda$K$^{+}$ and $\rm u\overline{u}$ in $\Lambda$K$^{-}$) or from different net strangeness for each system (S=0 for $\Lambda$K$^{+}$, and S=$-2$ for $\Lambda$K$^{-}$).
Finally, the $\Lambda$K systems exhibit source radii larger than expected from extrapolation from identical particle femtoscopic studies.
This effect is interpreted as resulting from the separation in space--time of the single-particle $\Lambda$ and K source distributions.
\end{abstract}
\end{titlepage}
\setcounter{page}{2}

%************************************************************************************************************************
%************************************************************************************************************************
\section{Introduction}
\label{sec:Introduction}

Femtoscopy is an experimental method used to study the space--time characteristics of the particle emitting sources in relativistic particle collisions~\cite{Lisa:2005dd, Kopylov:1972qw}.  
With this method, two- (or many-) particle relative-momentum correlation functions are used to connect the final-state momentum distributions to the space--time distributions of particle emission at freeze-out.  
The correlation functions are sensitive to quantum statistics, as well as strong and Coulomb final-state interactions (FSI).  
Current femtoscopic studies are able to extract the size, shape, and orientation of the pair emission regions, as well as offer estimates of the total time to reach kinetic decoupling and the duration of particle emission~\cite{Lisa:2005dd, Lisa:2008gf}.
The momentum and species dependence of femtoscopic measurements affirms the collective nature of the hot and dense matter created in heavy-ion collisions~\cite{Makhlin:1987gm, Akkelin:1995gh, Retiere:2003kf, Kisiel:2009eh}.
Non-identical particle analyses additionally allow for the measurement of the space--time separation of the single particle source regions~\cite{Lednicky:1995vk, Voloshin:1997jh, Retiere:2003kf}.

In addition to characterizing the source region, femtoscopy allows one to extract nuclear scattering parameters, many of which are difficult or impossible to measure otherwise.
The subjects of this analysis, \LamK pairs, interact only strongly; therefore, the studied femtoscopic signals are free from quantum statistical and Coulomb interaction effects.
Calculations within Quantum Chromodynamics (QCD), the theory of the strong interaction, are notoriously difficult except in the regime of weak coupling, where perturbative methods may be applied. 
The \LamK analysis presented here offers the possibility to access QCD measurements, which fall into the non-perturbative regime of QCD.
Furthermore, the \LamK scattering parameters were not previously known, and theoretical predictions are limited.
The extracted scattering parameters are compared to predictions obtained in the framework of chiral perturbation theory~\cite{Liu:2006xja,Mai:2009ce}.
Information about scattering parameters for similar systems are also very limited; past studies of kaon-proton scattering revealed the strong force is attractive in the K$^{-}$p interaction, and repulsive in that of the K$^{+}$p~\cite{Humphrey:1962zz, Hadjimichef:2002xe, Ikeda:2012au}.
Femtoscopy studies of K$^{-}$p and K$^{+}$p correlations carried out by ALICE allowed to constrain both interactions more precisely~\cite{PhysRevLett.124.092301}.

This paper presents the first measurements of the scattering parameters of \LamK pairs in all three charge combinations (\LamKchP, \LamKchM, and \LamKs).
The scattering parameters, along with pair emission source sizes, are extracted with a femtoscopic analysis of \LamK correlations in Pb--Pb collisions at $\sqrt{s_{\mathrm{NN}}}$ = 2.76 TeV measured by the ALICE experiment at the LHC.  
These correlations result from strong final-state interactions, and are fit with a parametrization by Lednick\'y and Lyuboshitz~\cite{Lednicky:82}.  
Extensive studies with the THERMINATOR 2~\cite{Chojnacki:2011hb} event generator are performed to account for both non-femtoscopic backgrounds as well as contributions from residual correlations induced by feed-down from particle decays.

The organization of this paper is as follows.  
In Sec.~\ref{sec:DataAnalysis} the data selection methods are briefly discussed.
In Sec.~\ref{sec:AnalysisMethods} the analysis techniques utilized in this study are presented.  
Here, the two-particle correlation function is introduced, as well as the theoretical models with which the data are fit.  
This section also includes descriptions of the handling of residual correlations, corrections accounting for finite track momentum resolution, treatment of the non-femtoscopic background, as well as a brief description of the systematic uncertainties estimation.  
The final results are presented in Sec.~\ref{sec:Results} and concluding remarks are given in Sec.~\ref{sec:Summary}.
Appendix~\ref{App:StavMethod} demonstrates an alternate approach to forming correlation functions, whose purpose here is to help eliminate the non-femtoscopic background.
Appendix~\ref{App:CoulombFitter} discusses the procedure needed to generate fit functions when both the strong and Coulomb interactions are present.
In Appendix~\ref{App:THERM}, the THERMINATOR 2 event generator is used to demonstrate the effect on a one-dimensional femtoscopic fit of a non-zero space-time separation between the single particle sources.
Throughout the text, the pair name is used as shorthand for the pair-conjugate system, which are found to be consistent (e.g., \LamKchP for \LamKchP $\oplus$ \ALamKchM, \LamKchM for \LamKchM $\oplus$ \ALamKchP, and \LamKs for \LamKs $\oplus$ \ALamKs), and \LamK is used to describe all \LamK combinations.

%************************************************************************************************************************
%************************************************************************************************************************
\section{Data analysis}
\label{sec:DataAnalysis}

This work reports on the analysis of Pb--Pb collisions at $\sqrt{s_{\mathrm{NN}}}$ = 2.76 TeV produced by the LHC and measured by the ALICE experiment~\cite{Aamodt:2008zz} in 2011.
Approximately 40 million events were analyzed, which were classified according to their centrality percentiles determined using the measured amplitudes in the V0 detectors~\cite{Abelev:2013qoq}.  
In order for an event to be included in the analysis, the position of the reconstructed event vertex must be within 10 cm of the center of the ALICE detector along the beam axis. 

Charged particle tracking was performed using the Time Projection Chamber (TPC)~\cite{2010NIMPA.622..316A} and the Inner Tracking System (ITS)~\cite{Abelevetal:2014dna}.  
The ITS allows for high spatial resolution of the primary (collision) vertex.
The momenta were determined by the tracking algorithm using tracks reconstructed with the TPC only and constrained to the primary vertex.
Tracks were selected from the central pseudorapidity region, $|\eta| < 0.8$.
A minimum requirement of 80 reconstructed TPC clusters was imposed, the purpose of which is to ensure both the quality of the track and good transverse momentum (\pt) resolution at large momenta, as well as to reject fake tracks.

Particle identification (PID) for reconstructed tracks was carried out using both the TPC and Time-Of-Flight (TOF) detectors~\cite{Abelev:2014ffa, Akindinov:2013tea}.  
For TPC PID, a parametrized Bethe-Bloch formula was used to calculate the specific energy loss $\langle \mathrm{d}E/\mathrm{d}x \rangle$ in the detector expected for a particle with a given mass and momentum.  
For TOF PID, the particle mass was determined using the time of flight as a function of track length and momentum.  
For each PID method, a value ($N_{\sigma}$) was assigned to each track denoting the number of standard deviations between the measured track and the expected signal at a given momentum for a particular hypothesis particle species.  
This procedure was applied for each track assuming four different particle species hypotheses--- electron, pion, kaon, and proton--- and for each hypothesis a different $N_{\sigma}$ value was obtained per detector.
These $N_{\sigma}$ values were used to identify primary \Kpm mesons, $\uppi$ and p daughters of \Lam hyperons, and $\uppi$ daughters of \Ks mesons, as well as to reject misidentified particles within each aforementioned sample.  

%************************************************************************************************************************
\subsection{K$^{\pm}$ selection}
\label{sec:KchSelection}
The single-particle selection criteria used to select charged kaon candidates are summarized in Table~\ref{tab:KchCuts}.
Track reconstruction for the charged kaons was performed using the TPC, and tracks within the range 0.14 $<$ \pt $<$ 1.5 GeV/$c$ were accepted for analysis.
To reduce the number of secondary particles (e.g., charged particles produced in the detector material, particles from weak decays, etc.) in the sample, a selection criterion is established based on the maximum distance-of-closest-approach (DCA) of the track to the primary vertex.
This is realized by imposing a restriction on the DCA in both the transverse and beam directions.

Particle identification was performed using both the TPC and TOF detectors via the $N_{\sigma}$ method. 
The $N_{\sigma}$ selection criteria become tighter with increasing momentum to reduce contamination within the samples, as the \Kpm signals begin to overlap more significantly with those from other particles, particularly e$^{\pm}$ and $\uppi^{\pm}$.
Rejection procedures are included to reduce the contamination in the \Kpm samples from electrons and pions.  
The specifics for the \Kpm selection are contained in Table~\ref{tab:KchCuts}.
The purity of the \Kpm collections, $P_{\mathrm{K}^{\pm}}$, was estimated to be approximately 97\% from a Monte-Carlo (MC) study based on HIJING~\cite{PhysRevD.44.3501} simulations using GEANT3~\cite{Brun:1082634} to model particle transport through the ALICE detectors. 
For a more detailed estimate of the \Kpm purity from an analysis employing similar methods, see~\cite{Acharya:2017qtq}.

\begin{table}[htbp]
 \centering
 \caption{Selection criteria for \Kpm mesons}
  \begin{tabular}{llll}
   \hline
   \hline 
   \multicolumn{4}{c}{\Kpm selection} \\
   \hline
   Transverse momentum $p_{\mathrm{T}}$ & 0.14 $< p_{\mathrm{T}} < 1.5$ GeV/\textit{c} \\
   $|\eta|$ & $< 0.8$ \\
   Transverse DCA to primary vertex & $<$ 2.4 cm \\
   Longitudinal DCA to primary vertex & $<$ 3.0 cm \\

   TPC and TOF $N_{\sigma}$ \\
   \quad $p <$ 0.4 GeV/\textit{c} & $N_{\sigma \mathrm{K,TPC}} <$ 2 \\
   \quad 0.4 $\leq p <$ 0.45 GeV/\textit{c} & $N_{\sigma \mathrm{K,TPC}} <$ 1 \\    
   \quad 0.45 $\leq p <$ 0.80 GeV/\textit{c} & $N_{\sigma \mathrm{K,TPC}} <$ 3 \\ 
   {} & $N_{\sigma \mathrm{K,TOF}} <$ 2 \\
   \quad 0.80 $\leq p <$ 1.0 GeV/\textit{c} & $N_{\sigma \mathrm{K,TPC}} <$ 3 \\
   {} & $N_{\sigma \mathrm{K,TOF}} <$ 1.5 \\  
   \quad $p \geq$ 1.0 GeV/\textit{c} & $N_{\sigma \mathrm{K,TPC}} <$ 3 \\
   {} & $N_{\sigma \mathrm{K,TOF}} <$ 1 \\  
   \hline
   
   Electron rejection: reject if all satisfied \\
   \quad $N_{\sigma\mathrm{e},\mathrm{TPC}} < $ 3 \\
   \quad $N_{\sigma\mathrm{e},\mathrm{TPC}} < N_{\sigma\mathrm{K},\mathrm{TPC}}$ \\
   \quad $N_{\sigma\mathrm{e},\mathrm{TOF}} < N_{\sigma\mathrm{K},\mathrm{TOF}}$ \\
   \hline
   
   Pion rejection:  reject if: \\
   \quad $p <$ 0.65 GeV/\textit{c} & TOF and TPC available & {} & $N_{\sigma \uppi,\mathrm{TPC}} <$ 3 \\
   {} & {} & {} & $N_{\sigma \uppi,\mathrm{TOF}} <$ 3 \\

   {} & Only TPC available & $p <$ 0.5 GeV/\textit{c} & $N_{\sigma \uppi,\mathrm{TPC}} <$ 3 \\
   {} & {} & 0.5 $\leq p <$ 0.65 GeV/\textit{c} & $N_{\sigma \uppi,\mathrm{TPC}} <$ 2 \\
   \quad 0.65 $\leq p <$ 1.5 GeV/\textit{c} & $N_{\sigma \uppi,\mathrm{TPC}} <$ 5 \\
   {} & $N_{\sigma \uppi,\mathrm{TOF}} <$ 3 \\
   \quad $p \geq$ 1.5 GeV/\textit{c} & $N_{\sigma \uppi,\mathrm{TPC}} <$ 5 \\
   {} & $N_{\sigma \uppi,\mathrm{TOF}} <$ 2 \\
   \hline
   \hline
  \end{tabular}
 \label{tab:KchCuts} 
\end{table}

%************************************************************************************************************************
\subsection{$\mathbf{K^{0}_{S}}$ and $\boldsymbol{\Lambda}$ selection}
\label{sec:V0Selection}

Electrically neutral \LamALam and \Ks particles are reconstructed through their weak decays: \Lam $\rightarrow$ p$\uppi^{-}$ (\ALam $\rightarrow \uppi^{+}\overline{\mathrm{p}}$) and \Ks $\rightarrow$ $\uppi^{+}\uppi^{-}$, with branching ratios 63.9\% and 69.2\%~\cite{PhysRevD.98.030001}, respectively.
The obtained candidates are denominated as \Vz particles due to their decay topology.
The selection criteria used are shown in Tables~\ref{tab:LamCuts} and~\ref{tab:K0sCuts}.
Aside from kinematic and PID selection methods (using TPC and TOF detectors), the tracks of the decay products (called \textit{daughters}) must also meet a minimum requirement on their impact parameter with respect to the primary vertex.  
The decay vertex of the \Vz is calculated based on the positions in which the two daughter tracks were closest.
To help in reducing combinatorial background, a maximum value is demanded on the distance of closest approach between the daughters (DCA \Vz daughters).
The positive and negative daughter tracks are combined to form the \Vz candidate, the momentum of which is the sum of the momenta of the daughters calculated in the condition in which they were closest to one another.

To select primary candidates, the impact parameter with respect to the primary vertex is used as a selection criterion for each \Vz.
Furthermore, a restriction is imposed on the pointing angle, $\theta_{\mathrm{PA}}$, between the \Vz momentum and the vector pointing from the primary vertex to the secondary \Vz decay vertex, which is achieved by appointing a minimum value on $\cos(\theta_{\mathrm{PA}})$ (``Cosine of pointing angle'' in Tables~\ref{tab:LamCuts} and~\ref{tab:K0sCuts}).

In order to remove the contamination to the \LamALam and \Ks samples due to misidentification of the protons and pions for each \Vz, the mass assuming different identities (\Lam, \ALam, and \Ks hypotheses) is calculated and utilized in a misidentification procedure.
The \Ks hypothesis ($m_{\mathrm{inv,~ K^{0}_{S}~ hyp.}}$) is calculated assuming $\uppi^{+}\uppi^{-}$ daughters, the \Lam hypothesis ($m_{\mathrm{inv,~ \Lambda~ hyp.}}$) assumes p$\uppi^{-}$ daughters, and the \ALam hypothesis ($m_{\mathrm{inv,~ \overline{\Lambda}~ hyp.}}$) assumes $\overline{\mathrm{p}}\uppi^{+}$ daughters. 
In the misidentification methods, the calculated masses are compared to the corresponding particle masses of the \Ks and \LamALam, $m_{\mathrm{PDG,\,K^{0}_{S}}}$ and $m_{\mathrm{PDG,\,\Lambda(\overline{\Lambda})}}$ respectively, as recorded by the Particle Data Group~\cite{PhysRevD.98.030001}.
For \LamALam selection, a candidate is concluded to be misidentified and is rejected if all of the following criteria are satisfied:

\begin{enumerate}
 \item $\left|m_{\mathrm{inv,\,K^{0}_{S}\,hyp.}} - m_{\mathrm{PDG,\,K^{0}_{S}}}\right| < $ 9.0 MeV/$c^{2}$,
 \item daughter particles pass daughter selection criteria intended for \Ks reconstruction,
 \item $\left|m_{\mathrm{inv,\,K^{0}_{S}\,hyp.}} - m_{\mathrm{PDG,\,K^{0}_{S}}}\right|~ < ~\left|m_{\mathrm{inv,\,\Lambda(\overline{\Lambda})\,hyp.}} - m_{\mathrm{PDG,\,\Lambda(\overline{\Lambda})}}\right|$.
\end{enumerate} 
Similarly, for \Ks selection, a candidate is rejected if all of the following criteria are satisfied for the \Lam case, or for the \ALam case:
\begin{enumerate}
 \item $\left|m_{\mathrm{inv},\,\Lambda(\overline{\Lambda})\,\mathrm{hyp.}} - m_{\mathrm{PDG},\,\Lambda(\overline{\Lambda})}\right| < $ 9.0 MeV/$c^{2}$,
 \item daughter particles pass daughter selection criteria intended for \LamALam reconstruction,
 \item $\left|m_{\mathrm{inv},\,\Lambda(\overline{\Lambda})\,\mathrm{hyp.}} - m_{\mathrm{PDG},\,\Lambda(\overline{\Lambda})}\right|~ < ~\left|m_{\mathrm{inv},\,\mathrm{K}^{0}_{S}\,\mathrm{hyp.}} - m_{\mathrm{PDG},\,\mathrm{K}^{0}_{S}}\right|$.
\end{enumerate} 

\begin{figure}[htp]
  \centering
  %%----start of first subfigure---
  \subfigure{
    \label{fig:Purity:a}
    \includegraphics[width=0.49\linewidth]{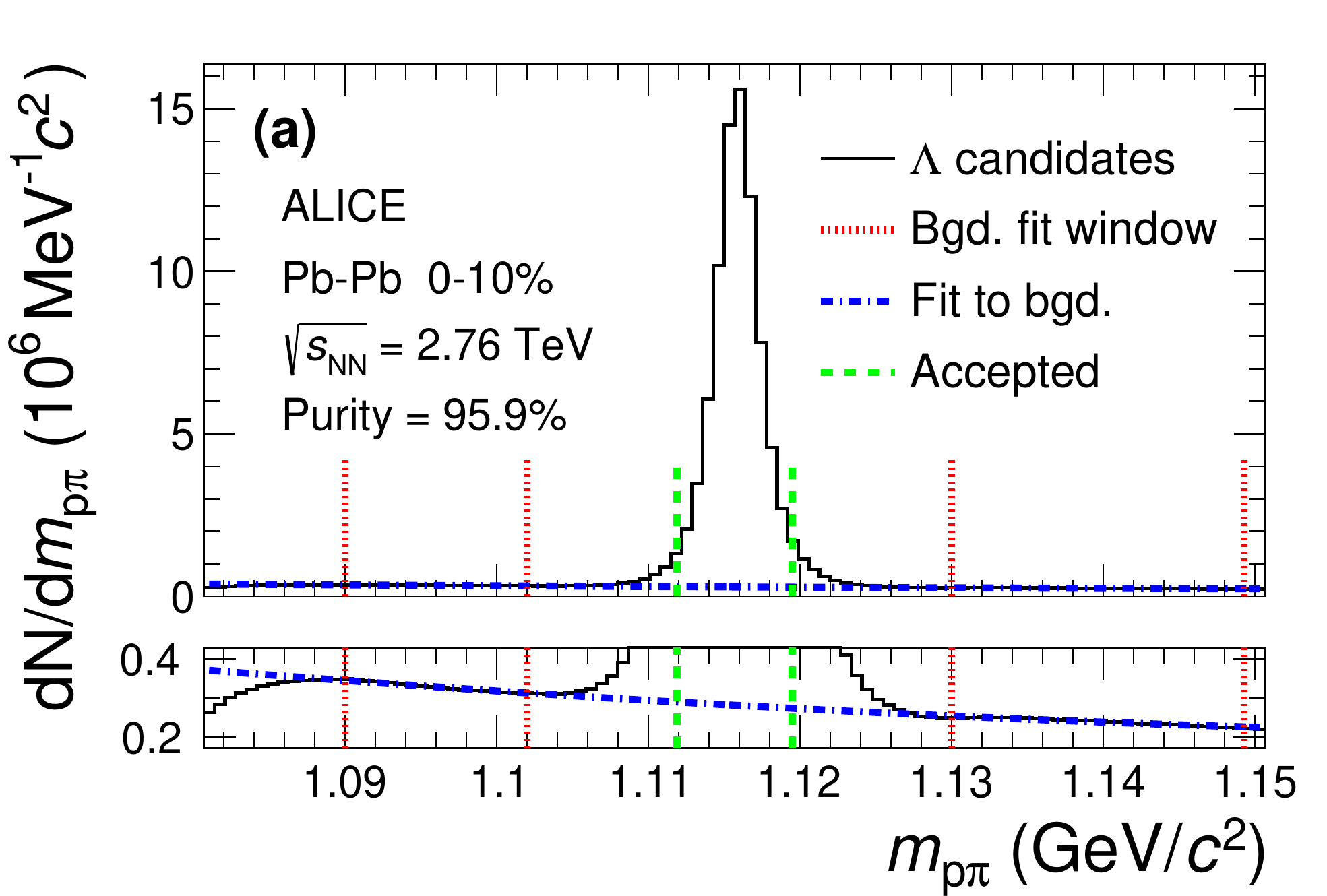}}
  %%----start of second subfigure---  
  \subfigure{
    \label{fig:Purity:b}
    \includegraphics[width=0.49\linewidth]{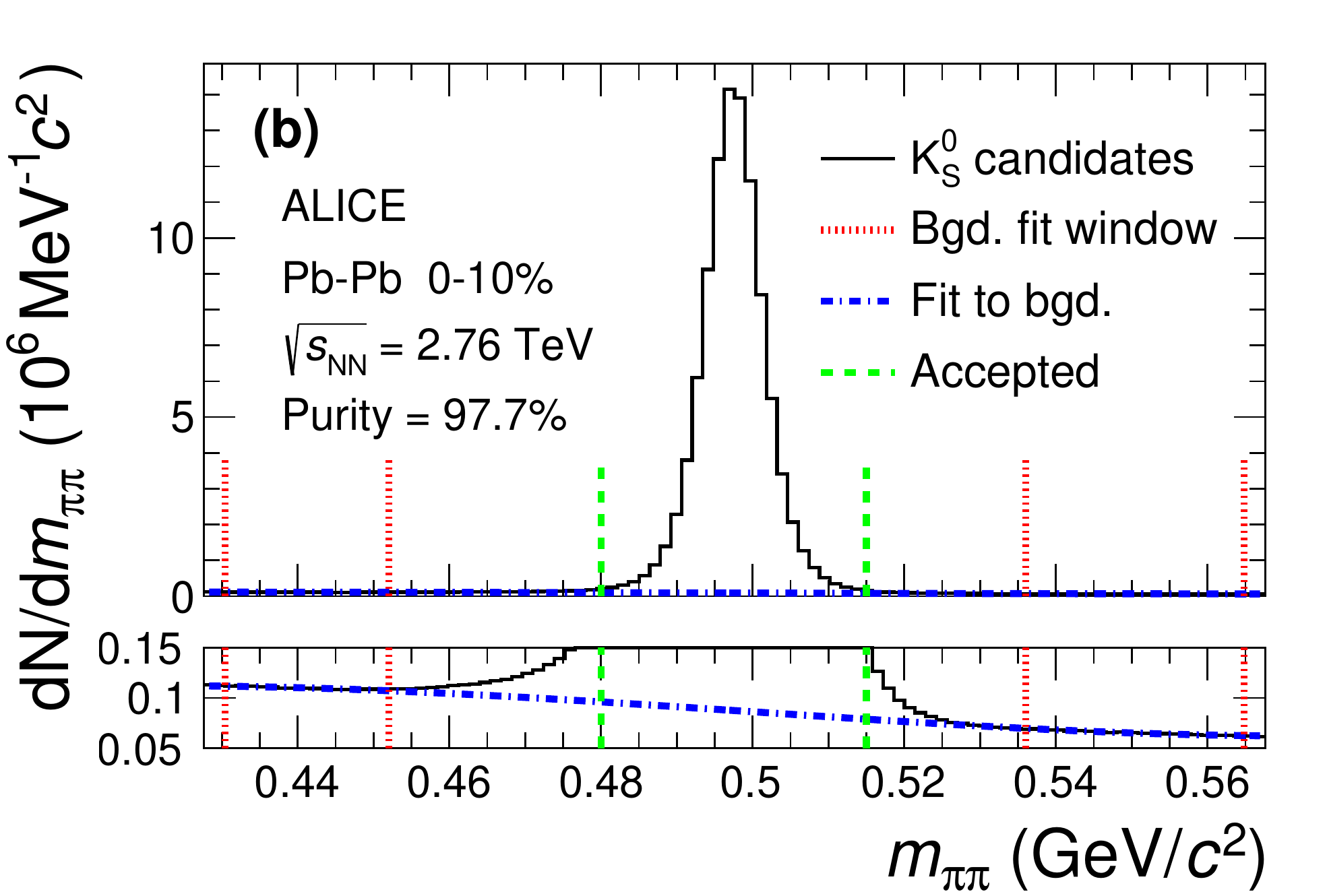}}
  %%----overall caption----
  \caption{
  (Color online) Invariant mass distributions in the 0--10\% centrality interval of (a) p$\uppi^{-}$ pairs showing the \Lam peak, and of (b) $\uppi^{+}\uppi^{-}$ pairs showing the \Ks peak, for \Vz candidates.  
  The bottom panels are zoomed to show the background with fit.  
  The vertical dashed (green) lines represent the selection restrictions used in the analyses, the vertical dotted (red) lines delineate the region over which the background was fit, and the dash-dotted (blue) line shows the background fit.
  }  
  \label{fig:Purity}
\end{figure}

A final restriction on the invariant mass (\minv) is applied to enhance the purity.
These selection criteria are shown in Tables~\ref{tab:LamCuts} and~\ref{tab:K0sCuts}.
To avoid any auto-correlation effects, all \Vz candidates within each single-particle collection (\Lam, \ALam, and \Ks separately) are ensured to have unique daughters. 
If a daughter is found to be shared among \Vz candidates in a given collection, only that with the smallest DCA to the primary vertex is kept.
This procedure ensures unique single-particle collections before particle pairs are constructed; the elimination of shared daughters between the particles within each pair is described below in Sec.~\ref{PairConstruction}.
The resulting invariant mass distributions for \Lam and \Ks collections in the 0--10\% centrality interval are shown in Fig.~\ref{fig:Purity}.
For the purity estimations, the background signal is extracted by fitting the \minv distribution with a fourth-order polynomial outside of the mass peak and assuming the distribution to continue smoothly beneath the mass peak.
The \Lam and \ALam purities are estimated to be $P_{\Lambda(\overline{\Lambda})} \approx 96\%$, and that of the \Ks is $P_{\mathrm{K^{0}_{S}}} \approx 98\%$.

\begin{table}[htbp]
 \centering 
 \caption{Selection criteria for \Lam and \ALam hyperons}
  \begin{tabular}{lll}
   \hline
   \hline  
   \multicolumn{3}{c}{\Lam [$\overline{\Lambda}$] selection} \\
   \hline
   Transverse momentum $p_{\mathrm{T}}$ & $> 0.4$ GeV/\textit{c} \\
   $|\eta|$ & $< 0.8$ \\
   Invariant mass & $|m_{\mathrm{\mathrm{p}\uppi}} - m_{\mathrm{PDG}}| < 3.8$ MeV/$c^{2}$ \\ 
   DCA to primary vertex & $<$ 0.5 cm \\
   Cosine of pointing angle & $>$ 0.9993 \\
   Decay length & $<$ 60 cm \\
   \hline
      
   \multicolumn{3}{c}{$\uppi$ and p daughter criteria} \\
   $|\eta|$ &  $< 0.8$ \\
   DCA $\uppi$p daughters & $<$ 0.4 cm \\
   \hline
     
   \multicolumn{3}{c}{$\uppi$-specific} \\
   $p_{\mathrm{T}}$ & $> 0.16$ GeV/\textit{c} \\
   DCA to primary vertex & $>$ 0.3 cm \\
   TPC and TOF $N_{\sigma}$ \\
   \quad $p <$ 0.5 GeV/\textit{c} & $N_{\sigma, \mathrm{TPC}} <$ 3 \\
   \quad $p \geq$ 0.5 GeV/\textit{c} &  TOF \& TPC available & $N_{\sigma, \mathrm{TPC}} <$ 3 \\
   {} & {} & $N_{\sigma, \mathrm{TOF}} <$ 3 \\
   {} & Only TPC available & $N_{\sigma, \mathrm{TPC}} <$ 3 \\
   \hline

   \multicolumn{3}{c}{p-specific} \\
   $p_{\mathrm{T}}$ & $ > $ 0.5(p) [0.3($\overline{\mathrm{p}}$)] GeV/\textit{c} \\
   DCA to primary vertex & $>$ 0.1 cm \\
   TPC and TOF $N_{\sigma}$ \\
   \quad $p <$ 0.8 GeV/\textit{c} & $N_{\sigma, \mathrm{TPC}} <$ 3 \\
   \quad $p \geq$ 0.8 GeV/\textit{c} &  TOF \& TPC available & $N_{\sigma, \mathrm{TPC}} <$ 3 \\
   {} & {} & $N_{\sigma, \mathrm{TOF}} <$ 3 \\
   {} & Only TPC available & $N_{\sigma, \mathrm{TPC}} <$ 3 \\
   \hline
   \hline 
  \end{tabular}
 \label{tab:LamCuts} 
\end{table}

\begin{table}[htbp]
 \centering
 \caption{Selection criteria for \Ks mesons}
  \begin{tabular}{lll}
   \hline
   \hline
   \multicolumn{3}{c}{\Ks selection} \\
   \hline
   Transverse momentum $p_{\mathrm{T}}$ & $> 0.2$ GeV/\textit{c} \\
   $|\eta|$ & $< 0.8$ \\
   Invariant mass & $0.480 < m_{\mathrm{\uppi^{+}\uppi^{-}}} < 0.515$ GeV/$c^{2}$ \\
   DCA to primary vertex & $<$ 0.3 cm \\
   Cosine of pointing angle & $>$ 0.9993 \\
   Decay length & $<$ 30 cm \\
   \hline
        
   \multicolumn{3}{c}{$\uppi^{\pm}$ daughter criteria} \\
   $p_{\mathrm{T}}$ & $>$ 0.15 GeV/\textit{c} \\
   $|\eta|$ &  $< 0.8$ \\
   DCA $\uppi^{+}\uppi^{-}$ daughters & $<$ 0.3 cm \\
   DCA to primary vertex & $>$ 0.3 cm \\
   TPC and TOF $N_{\sigma}$ \\
   \quad $p <$ 0.5 GeV/\textit{c} & $N_{\sigma, \mathrm{TPC}} <$ 3 \\
   \quad $p \geq$ 0.5 GeV/\textit{c} &  TOF \& TPC available & $N_{\sigma, \mathrm{TPC}} <$ 3 \\
   {} & {} & $N_{\sigma, \mathrm{TOF}} <$ 3 \\
   {} & Only TPC available & $N_{\sigma, \mathrm{TPC}} <$ 3 \\
   \hline
   \hline
  \end{tabular}
 \label{tab:K0sCuts} 
\end{table}

%************************************************************************************************************************
\subsection{Pair construction}
\label{PairConstruction}

In order to reduce the contamination to the two-particle correlations due to pairs sharing daughters, track splitting (two tracks reconstructed from one particle), and track merging (one track reconstructed from two particles), two main pair rejection procedures are applied: a shared daughter restriction, and an average separation constraint.
The purpose of the shared daughter restriction is to ensure the first particle in the pair is unique from the second.  
For pairs formed of two V$^{0}$s (i.e., \LamKs), this is implemented by removing all pairs which share a daughter.  
For a pair formed of a single \Vz and a charged track (i.e., \LamKpm), the restriction removes all pairs in which the charged track is also claimed as a daughter of the \Vz.

The purpose of the average separation constraint is to remove splitting and merging effects, and it is employed in the following way.  
The average separation between two tracks is calculated using their spatial separation as determined at several points throughout the TPC (every 20 cm radially from 85 cm to 245 cm).
For the \LamKs analysis, which involves two \Vz particles, a minimum average separation constraint of 6 cm between the like-charge daughters in the pairs is imposed (for example, between the p daughter of the \Lam and the $\uppi^{+}$ daughter of the \Ks).
For the \LamKpm analyses, a minimum average separation constraint of 8 cm is enforced between the \Kpm and the \Lam daughter sharing the same charge (for example, in the \LamKchP analysis, between the p daughter of the \Lam and the \KchP).
Splitting and merging effects between oppositely charged tracks were found to be negligible, therefore no constraints on unlike-charge tracks are imposed.

%************************************************************************************************************************
%************************************************************************************************************************
\section{Analysis methods}
\label{sec:AnalysisMethods}

%************************************************************************************************************************
\subsection{Correlation function}
\label{sec:CorrelationFunction}
The correlation function for particles $a$ and $b$, $C_{ab}(\mathbf{p}_{a},\mathbf{p}_{b})$, is defined as the ratio of the probability of simultaneously measuring two particles with momenta $\mathbf{p}_{a}$ and $\mathbf{p}_{b}$, to the product of the single-particle probabilities.
These probabilities are directly related to the covariant two-particle spectrum, $E_{a}E_{b}\frac{\mathrm{d}^{6}N_{ab}}{\mathrm{d}^{3}p_{a}\mathrm{d}^{3}p_{b}}$, and the single-particle spectra, $E_{a(b)}\frac{\mathrm{d}^{3}N_{a(b)}}{\mathrm{d}^{3}p_{a(b)}}$, and the correlation function may be written
\begin{equation}
  C_{ab}(\mathbf{p}_{a},\mathbf{p}_{b}) = \frac{E_{a}E_{b}\frac{\mathrm{d}^{6}N_{ab}}{\mathrm{d}^{3}p_{a}\mathrm{d}^{3}p_{b}}}{\big( E_{a}\frac{\mathrm{d}^{3}N_{a}}{\mathrm{d}^{3}p_{a}} \big) \big( E_{b}\frac{\mathrm{d}^{3}N_{b}}{\mathrm{d}^{3}p_{b}} \big)},
\label{eqn:CfRatioSpectra}
\end{equation}
where $N_{ab}$ is the yield of particle pairs, $E_{a(b)}$ is the energy, $p_{a(b)}$ is the three-momentum, and $N_{a(b)}$ is the yield of particles $a(b)$.
Theoretically, the correlation function may be expressed as in the Koonin-Pratt equation~\cite{Koonin:1977fh, Pratt:1990zq},
\begin{equation}
 C(\mathbf{k^{*}}) = \int S_{\mathbf{P}}(\mathbf{r^{*}})|\Psi_{\mathbf{k^{*}}}(\mathbf{r^{*}})|^{2}\mathrm{d}^{3}r^{*},
\label{eqn:KooninPrattEqn}
\end{equation}
where $\mathbf{k}^{*}$ is the relative momentum of the pair (defined as $\mathbf{k}^{*} = \frac{1}{2}|\mathbf{p}_{a}^{*}-\mathbf{p}_{b}^{*}|$, where $\mathbf{p}_{a}^{*}$ and $\mathbf{p}_{b}^{*}$ are the momenta of the two particles) in the pair rest frame (PRF, denoted with an asterisk $^{*}$), $\mathbf{r}^{*}$ is the relative separation, $S_{\mathbf{P}}(\mathbf{r^{*}})$ is the pair source distribution, and $\Psi_{\mathbf{k^{*}}}(\mathbf{r^{*}})$ is the two-particle wave-function.

In practice, the correlation function is formed experimentally as
\begin{equation}
  C(k^{*}) = \mathcal{N}\frac{A(k^{*})}{B(k^{*})},
\label{eqn:CfExp}
\end{equation}
where $A(k^{*})$ is the signal distribution, $B(k^{*})$ is the reference distribution, and $\mathcal{N}$ is a normalization parameter.  
The reference distribution is used to correct for the phase-space effects, leaving only the physical effects in the correlation function. 
The normalization parameter is chosen such that the mean value of the correlation function equals unity for \kstar $\in$ [0.32, 0.4] GeV/$c$.
The signal distribution is the same-event distribution of particle pairs.
The reference distribution, $B(k^{*})$, is obtained using mixed-event pairs~\cite{Kopylov:1974th}, i.e., particles from a given event are paired with those from another event.
For this analysis, each event is combined with five others for the reference distribution construction.
To be included in the mixing pool, an event must contain at least one particle of each type from the pair of interest (e.g., for the \LamKs analysis, an accepted event must contain at least one \Lam and at least one \Ks).
In order to mix similar events, only those of like centrality (within 5\%) and of like primary vertex position (within 2 cm) are combined.

This analysis presents correlation functions for three centrality percentile ranges (0--10\%, 10--30\%, and 30--50\%), and is integrated in pair transverse momentum ($k_{\mathrm{T}} = \frac{1}{2}|\mathbf{p}_{\mathrm{T,1}}+\mathbf{p}_{\mathrm{T,2}}|$) due to limited data.
The $k_{\mathrm{T}}$ dependences of the three \LamK charge combinations should be comparable, so an integrated analysis is acceptable.

%************************************************************************************************************************
\subsection{Modeling the correlation function}
\label{sec:ModelingCF}

In the absence of the Coulomb interaction, the correlation function can be described analytically with a model derived by Lednick\'y and Lyuboshitz~\cite{Lednicky:82}.
Within the model, the (non-symmetrized) two-particle wave function is expressed as a superposition of a plane wave and diverging spherical wave, and the complex scattering amplitude, $f^{S}(k^{*})$, is evaluated via the effective range approximation,
\begin{equation}
\begin{aligned}
f^{S}(k^{*}) = \left( \frac{1}{f^{S}_{0}} + \frac{1}{2}d^{S}_{0}k^{*2} - ik^{*} \right)^{-1},
\end{aligned}
\label{eqn:ScatteringParam}
\end{equation}
where $f^{S}_{0}$ is the complex s-wave scattering length, $d^{S}_{0}$ is the effective range of the interaction, and $S$ denotes the total spin of the particular pair.
The sign convention is such that a positive real component of the scattering length, $\Re f_{0}$, represents an attractive interaction, while a negative value represents a repulsion.
A spherically symmetric Gaussian distribution with radius $R_{\mathrm{inv}}$  is assumed for the pair emission source in the PRF.
With these assumptions, utilizing the Koonin-Pratt equation (Eq.~(\ref{eqn:KooninPrattEqn})), the correlation function for non-identical particle pairs is modeled by~\cite{Lednicky:82}
\begin{equation}
\begin{aligned}
C(k^{*})_{\mathrm{Lednick\acute{y}}} = &1 + \sum_{S}\rho_{S}\left[\frac{1}{2}\left|\frac{f^{S}(k^{*})}{R_{\mathrm{inv}}}\right|^2\left(1-\frac{d^{S}_{0}}{2\sqrt{\pi}R_{\mathrm{inv}}}\right) \right. \\
&+ \left. \frac{2\Re f^{S}(k^{*})}{\sqrt{\pi}R_{\mathrm{inv}}}F_{1}(2k^{*}R_{\mathrm{inv}})-\frac{\Im f^{S}(k^{*})}{R_{\mathrm{inv}}}F_{2}(2k^{*}R_{\mathrm{inv}})\right],
\end{aligned}  
\label{eqn:LednickyEqn}
\end{equation}
where $\Re f^{S}(k^{*})$ and $\Im f^{S}(k^{*})$ denote the real and imaginary parts of the complex scattering length, respectively, and $F_{1}$ and $F_{2}$ are analytic functions~\cite{Lednicky:82}.
The weight factor, $\rho_{S}$, is the normalized emission probability for a state of total spin $S$; in the assumed case of unpolarized emission, $\rho_{S} = (2S+1)/[(2j_{1}+1)(2j_{2}+1)]$, where $j_{1,2}$ are the spins of the particles in the pair.
The \Lam hyperon is spin-1/2 and K mesons are spin-0, so the \LamK system only has one possible total spin state $S$, and therefore $C(k^{*})$ in Eq.~(\ref{eqn:LednickyEqn}) has only a single term.
In the following, the $S$ superscript is dropped from all scattering parameters.

%************************************************************************************************************************
\subsection{Residual correlations}
\label{ResidualCorrelations}

The purpose of this analysis is to study the interaction and scale of the emitting source of the primary \LamK pairs.
However, in practice some of the selected particles originate as products from other decaying particles after kinetic freeze-out (secondary particles), and some of the final pairs contain a misidentified member.
In both cases, these contribute to the observed correlation function, and obscure its relation to the primary \LamK system.
The net contribution from fake pairs, which contain at least one misidentified member, is taken to average to unity, in which case they simply attenuate the femtoscopic signal.
Pairs in which at least one member originates from a particle decay (e.g., \LamKchP from $\Sigma^{0}$\KchP) carry information about the parent system.
In effect, the correlation between the parents will be visible, although smeared out, in the daughters' signal.
This is termed a residual correlation resulting from feed-down.  
As described in the following, the main sources of residual correlations in the \LamK systems result from \Lam hyperons which have been produced from $\Sigma^{0}$, $\Xi^{0}$, and $\Xi^{-}$ decays. 

The measured correlation function is a combination of the genuine \LamK correlation with contributions from particle decays and impurities~\cite{Kisiel:2014mma},
\begin{equation}
\begin{aligned}
\label{eqn:CfwRes} 
 C_{\mathrm{measured}}(k^{*}_{\Lambda\mathrm{K}}) &= 1 + \sum\limits_{ij}  \lambda'_{ij}[C_{ij}(k^{*}_{\Lambda\mathrm{K}})-1], \\
\end{aligned} 
\end{equation}
with
\begin{equation}
\begin{aligned}
\label{eqn:CfwRes2} 
 \lambda_{ij}' &= \lambda_{\mathrm{Fit}}\lambda_{ij}, \\
 \sum\limits_{ij}\lambda_{ij}' &=  \lambda_{\mathrm{Fit}}\sum\limits_{ij}\lambda_{ij} = \lambda_{\mathrm{Fit}},
\end{aligned} 
\end{equation}
where the $ij$ terms include the primary \LamK contribution together with the contributions from residual feed-down and impurities.
More specifically, $C_{ij}(k^{*}_{\Lambda\mathrm{K}})$ is the correlation function of the parent system expressed in terms of the relative momentum of the daughter \LamK pair.  
The $\lambda_{ij}$ parameters serve as weights dictating the relative strength of each component's contribution to the observed signal, and are normalized to unity (i.e., $\sum_{ij} \lambda_{ij} = 1$, where $ij$ includes also the primary \LamK component)~\cite{Kisiel:2014mma, Acharya:2018gyz}.
When the experimental correlation functions are fit, the individual $\lambda_{ij}$ are fixed (and whose values can be found in Table~\ref{tab:LambdaValues_3Res}), but the parameter $\lambda_{\mathrm{Fit}}$ in Eq.~(\ref{eqn:CfwRes2}) is left free.

To model the parent correlation function expressed in the relative momentum of the daughter pair, a transform matrix is utilized,
\begin{equation}
  C_{ij}(k^{*}_{\Lambda\mathrm{K}}) \equiv \frac{\sum\limits_{k^{*}_{ij}} C_{ij}\left(k^{*}_{ij}\right) T\left(k^{*}_{ij},k^{*}_{\Lambda\mathrm{K}}\right)}{\sum\limits_{k^{*}_{ij}} T\left(k^{*}_{ij},k^{*}_{\Lambda\mathrm{K}}\right)},
\label{eqn:ResidualsTransform}
\end{equation}
where $T(k^{*}_{ij},k^{*}_{\Lambda\mathrm{K}})$ is the transform matrix, which is generated with the THERMINATOR 2~\cite{Chojnacki:2011hb} simulation. 
The transform matrix describes the decay kinematics of the parent system into the daughter system, and is essentially an unnormalized probability distribution mapping the \kstar of the parent pair to that of the daughter pair when one or both parents decay (see~\cite{Kisiel:2014mma} for more details).

The contribution of a parent system (e.g., $\Sigma^{0}$\KchP) to the daughter correlation function (e.g., \LamKchP) in the fit procedure is determined by modeling the parent system's correlation function and running it through the appropriate transform matrix.
Since the interactions between these particles are not known, some assumptions must be made.
When modeling the parent systems, the source radii are assumed to be equal to those of the daughter \LamK systems.
Furthermore, Coulomb-neutral parent pairs are assumed to share the same scattering parameters as the \LamK daughter pair, and the parent correlation function is modeled using Eq.~(\ref{eqn:LednickyEqn}).
During the fit process, these source radii and scattering parameters are left free, as described in Sec.~\ref{SummarizedFitProcedure}.
For the \XiKpm parent system, where the constituents interact via both the strong and Coulomb interactions, no analytical expression exists to model the correlation function (see App.~\ref{App:CoulombFitter}), and the experimental \XiKpm data are used.
However, the \XiKpm correlation function is dominated by the contribution from the Coulomb interaction, and may be sufficiently modeled with a Coulomb-only scenario (in which the strong interaction is assumed to be negligible) for this analysis, to yield consistent results.

The $\lambda_{ij}$ parameters dictate the relative strength of each contribution to the correlation function, and can be estimated using the THERMINATOR 2 and HIJING simulations.
More specifically, a $\lambda_{ij}$ parameter is estimated as the total number of \LamK pairs in the sample originating from source $ij$ ($N_{ij}$) divided by the total number of \LamK pairs.
For a given \LamK source, the number of detected pairs depends on both the raw yield and the reconstruction efficiency.
The relevant reconstruction efficiencies are those of the daughters under study, not of the parent particles; e.g., when determining the contribution of the \XiKchP system to the \LamKchP correlation function, the reconstruction efficiency of the $\Xi^{-}$ is not relevant, but that of the secondary \Lam originating from a $\Xi^{-}$ decay is. 
The reconstruction efficiencies ($RE_{ij}$) are estimated with HIJING simulations using GEANT3 to model particle transport through the detector.
HIJING events are generated from a superposition of PYTHIA pp collisions, and lack the strangeness saturation of a fully thermalized medium.
As a result, HIJING is unreliable in providing the yields needed for this analysis, and, instead, the yields are estimated with the THERMINATOR 2 simulation ($N_{ij}^{\scaleto{\mathrm{THERM}}{4pt}}$).
The number of \LamK pairs from source $ij$ is then estimated as the product of the yield with the associated reconstruction efficiency, $N_{ij} = N_{ij}^{\scaleto{\mathrm{THERM}}{4pt}}RE_{ij}^{\scaleto{\mathrm{HIJING}}{4pt}}$.
Finally, the $\lambda_{ij}$ are estimated as
\begin{equation}
\lambda_{ij} = \frac{N_{ij}}{N_{\mathrm{Total}}} = \frac{N_{ij}^{\scaleto{\mathrm{THERM}}{4pt}}RE_{ij}^{\scaleto{\mathrm{HIJING}}{4pt}}}{\sum\limits_{ij} N_{ij}^{\scaleto{\mathrm{THERM}}{4pt}}RE_{ij}^{\scaleto{\mathrm{HIJING}}{4pt}}}.
\end{equation}

Femtoscopic analyses are sensitive to the pair emission structure at kinetic freeze-out.
Therefore, within femtoscopy, any particle which originates from a particle decay before last rescattering is considered primary.
The THERMINATOR 2 simulation shows that the \Lam hyperons and K mesons decay from a large number of particle species ($\sim$50 \Lam parent species, and $\sim$70 K parent species), and the most significant contributing pair systems are $\Sigma^{0}$K, $\Xi^{-}$K, $\Xi^{0}$K, $\Sigma^{*+}$K, $\Sigma^{*-}$K, $\Sigma^{*0}$K, $\Lambda\mathrm{K}^{*}$, $\Sigma^{0}\mathrm{K}^{*}$, $\Xi^{-}\mathrm{K}^{*}$, and $\Xi^{0}\mathrm{K}^{*}$.
However, the simulation does not include a hadronic rescattering phase, and not all of the aforementioned pair systems will survive until kinetic freeze-out.
The systems resulting from electromagnetic or weak decays ($\Sigma^{0}$, $\Xi^{-}$, and $\Xi^{0}$) will survive long after kinetic freeze-out, and will contribute residual signals to the \LamK correlation functions.
The majority of the remaining contributors decay via the strong interaction with mean proper lifetimes less than a few fm/$c$, and their daughters should always be considered primary.
The mean proper lifetime of the parent is used to judge whether or not the daughter is treated as primary.
A decay product is considered primary if its parent has a mean proper lifetime $\tau$ satisfying $\tau < \tau_{\mathrm{max}}$, where $c\tau_{\mathrm{max}} =$ 10 fm for this analysis.
Changing $\tau_{\mathrm{max}}$ only moderately affects the $\lambda_{ij}$ parameters, and the effect is included in the estimation of the systematic uncertainties.
In order for a pair to be considered primary, both particles in the pair must be considered primary. 
If either parent has $\tau > \tau_{\mathrm{max}}$, the daughter pair contributes to the ``Other" category when calculating $\lambda_{ij}$ parameters.
For this mixture of pair systems, all with different two-particle interactions and single-particle source distributions, the net correlation effect is taken to average to unity.

Residual contributions from $\Sigma^{0}$, $\Xi^{0}$, $\Xi^{-}$ are accounted for in the fit.
The $\lambda_{ij}$ values used can be found in Table~\ref{tab:LambdaValues_3Res}, which also includes values for ``Other'' and ``Fakes''.  
The ``Other'' category contains pairs which are not considered primary, and which do not originate from the residual contributors accounted for in the fit.  
The ``Fakes'' category represents pairs that are mistakenly identified as \LamK.
The corresponding $\lambda_{\mathrm{Fakes}}$ is calculated as $\lambda_{\mathrm{Fakes}} = 1-PP_{\Lambda\mathrm{K}}$, where $PP_{\Lambda\mathrm{K}}$ is the \LamK pair purity, estimated as the product of the two single-particle purities ($PP_{\Lambda\mathrm{K}} = P_{\Lambda}P_{\mathrm{K}}$).
The correlations in both of these categories (``Other'' and ``Fakes'') are assumed to average to unity, and pairs in these categories therefore only contribute by attenuating the signal. 

\begin{table}[htbp] 
 \centering
 \caption{Weight parameters ($\lambda_{ij}$) for the individual components of the \LamK correlation functions}
 \begin{tabular}{cc c cc c cc c cc}
  \hline
  \hline
  \multicolumn{2}{c}{\LamKchP} & \multicolumn{1}{c}{} & \multicolumn{2}{c}{\ALamKchM} & \multicolumn{1}{c}{} & \multicolumn{2}{c}{\LamKchM} & \multicolumn{1}{c}{} & \multicolumn{2}{c}{\ALamKchP} \\
  \cline{1-2} \cline{4-5} \cline{7-8} \cline{10-11}
  Source & $\lambda$ value & \multicolumn{1}{c}{} & Source & $\lambda$ value & \multicolumn{1}{c}{} & Source & $\lambda$ value & \multicolumn{1}{c}{} & Source & $\lambda$ value \\
  \cline{1-2} \cline{4-5} \cline{7-8} \cline{10-11}
  Primary & 0.509 & \multicolumn{1}{c}{} & Primary & 0.509 & \multicolumn{1}{c}{} & Primary & 0.509 & \multicolumn{1}{c}{} & Primary & 0.510 \\
  $\Sigma^{0}$K$^{+}$ & 0.108 & \multicolumn{1}{c}{} & $\overline{\Sigma}^{0}$K$^{-}$ & 0.107 & \multicolumn{1}{c}{} & $\Sigma^{0}$K$^{-}$ & 0.107 & \multicolumn{1}{c}{} & $\overline{\Sigma}^{0}$K$^{+}$ & 0.108 \\  
  $\Xi^{0}$K$^{+}$ & 0.037 & \multicolumn{1}{c}{} & $\overline{\Xi}^{0}$K$^{-}$ & 0.034 & \multicolumn{1}{c}{} & $\Xi^{0}$K$^{-}$ & 0.037 & \multicolumn{1}{c}{} & $\overline{\Xi}^{0}$K$^{+}$ & 0.035 \\  
  $\Xi^{-}$K$^{+}$ & 0.048 & \multicolumn{1}{c}{} & $\overline{\Xi}^{+}$K$^{-}$ & 0.044 & \multicolumn{1}{c}{} & $\Xi^{-}$K$^{-}$ & 0.048 & \multicolumn{1}{c}{} & $\overline{\Xi}^{+}$K$^{+}$ & 0.045 \\  
  Other & 0.218 & \multicolumn{1}{c}{} & Other & 0.228 & \multicolumn{1}{c}{} & Other & 0.221 & \multicolumn{1}{c}{} & Other & 0.225 \\  
  Fakes & 0.079 & \multicolumn{1}{c}{} & Fakes & 0.079 & \multicolumn{1}{c}{} & Fakes & 0.079 & \multicolumn{1}{c}{} & Fakes & 0.079 \\
  
  \multicolumn{11}{c}{} \\

  \multicolumn{3}{c}{} & \multicolumn{2}{c}{\LamKs} & \multicolumn{1}{c}{} & \multicolumn{2}{c}{\ALamKs} & \multicolumn{3}{c}{} \\ 
  \cline{4-5} \cline{7-8}
  \multicolumn{3}{c}{} & Source & $\lambda$ value & \multicolumn{1}{c}{} & Source & \multicolumn{1}{c}{$\lambda$ value} & \multicolumn{3}{c}{} \\
  \cline{4-5} \cline{7-8}
  \multicolumn{3}{c}{} & Primary & 0.531 & \multicolumn{1}{c}{} & Primary & \multicolumn{1}{c}{0.532} & \multicolumn{3}{c}{} \\  
  \multicolumn{3}{c}{} & $\Sigma^{0}$K$^{0}_{\mathrm{S}}$ & 0.118 & \multicolumn{1}{c}{} & $\overline{\Sigma}^{0}$K$^{0}_{\mathrm{S}}$ & \multicolumn{1}{c}{0.118} & \multicolumn{3}{c}{} \\  
  \multicolumn{3}{c}{} & $\Xi^{0}$K$^{0}_{\mathrm{S}}$ & 0.041 & \multicolumn{1}{c}{} & $\overline{\Xi}^{0}$K$^{0}_{\mathrm{S}}$ & \multicolumn{1}{c}{0.038} & \multicolumn{3}{c}{} \\  
  \multicolumn{3}{c}{} & $\Xi^{-}$K$^{0}_{\mathrm{S}}$ & 0.053 & \multicolumn{1}{c}{} & $\overline{\Xi}^{+}$K$^{0}_{\mathrm{S}}$ & \multicolumn{1}{c}{0.049} & \multicolumn{3}{c}{} \\  
  \multicolumn{3}{c}{} & Other & 0.189 & \multicolumn{1}{c}{} & Other & \multicolumn{1}{c}{0.195} & \multicolumn{3}{c}{} \\  
  \multicolumn{3}{c}{} & Fakes & 0.069 & \multicolumn{1}{c}{} & Fakes & \multicolumn{1}{c}{0.069} & \multicolumn{3}{c}{} \\
  \hline
  \hline
 \end{tabular}
 \label{tab:LambdaValues_3Res}
\end{table}

%************************************************************************************************************************
\subsection{Momentum resolution corrections}
\label{MomentumResolutionCorrections}

Finite track momentum resolution causes the reconstructed relative momentum (\krec) of a pair to differ from the true value (\ktrue).
This is accounted for through the use of a response matrix generated with HIJING simulations.
With this approach, the resolution correction is applied on-the-fly during the fitting process by propagating the theoretical (fit) correlation function through the response matrix, according to
\begin{equation}
  C(k^{*}_{\mathrm{Rec}}) = \dfrac{\sum\limits_{k^{*}_{\mathrm{True}}}M_{k^{*}_{\mathrm{Rec}},k^{*}_{\mathrm{True}}}C(k^{*}_{\mathrm{True}})}{\sum\limits_{k^{*}_{\mathrm{True}}}M_{k^{*}_{\mathrm{Rec}},k^{*}_{\mathrm{True}}}},
\label{eqn:MomResCorrection}
\end{equation}
where $M_{k^{*}_{\mathrm{Rec}},k^{*}_{\mathrm{True}}}$ is the response matrix, $C(k^{*}_{\mathrm{True}})$ is the correlation as a function of \ktrue, and the denominator normalizes the result.

%************************************************************************************************************************
\subsection{Non-femtoscopic background}
\label{NonFlatBackground}

A significant non-femtoscopic background is observed in all of the studied \LamK correlations, which increases with decreasing centrality, is the same amongst all \LamKpm pairs, and is more pronounced in the \LamKs system (the difference in \LamKpm and \LamKs backgrounds is due mainly to a difference in kinematic selection criteria).  
The background is primarily due to particle collimation associated with elliptic flow, and results from mixing events with unlike event planes~\cite{Kisiel:2017}.
The effect produces the observed suppression at intermediate \kstar, and should also lead to an enhancement at low \kstar.
To best describe the experimental data, an understanding of the non-femtoscopic background is needed in the low \kstar femtoscopic signal region, but an isolated view of it is only possible outside of such a region.

\begin{figure}[h]
  \centering
  \includegraphics[width=\textwidth]{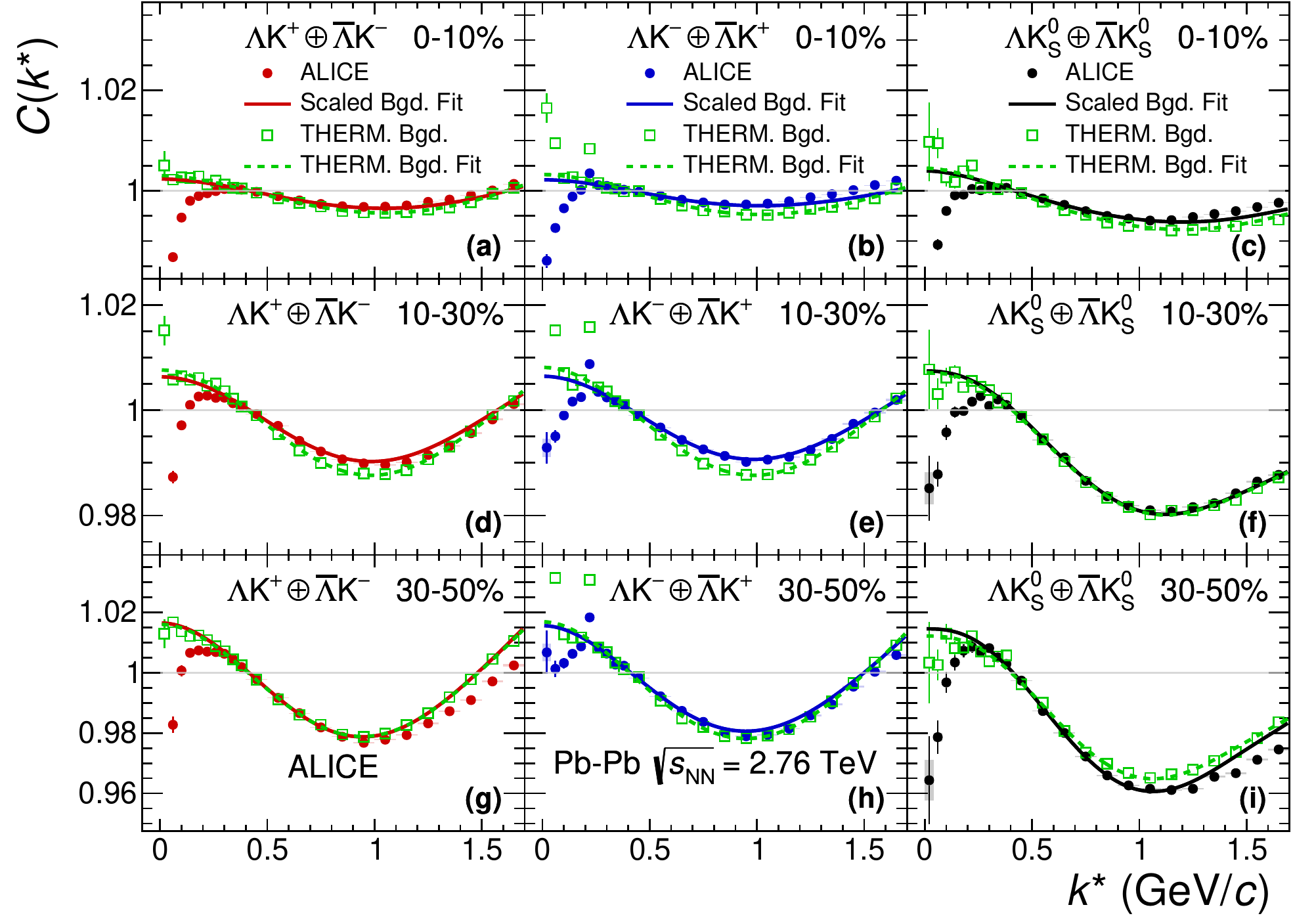}
  \caption[Backgrounds with THERMINATOR 2]
  {
  (Color online) THERMINATOR 2 simulation (open squares) together with experimental data (closed circles).
  Statistical (lines) and systematic (boxes) uncertainties are shown for the experimental data.
  Results are shown for \LamKchP (left), \LamKchM (middle), and \LamKs (right).
  Rows differentiate the different centrality intervals (0--10\% in the top, 10--30\% in the middle, and 30--50\% in the bottom).
  A sixth-order polynomial fit to the simulation is shown as a dashed curve.  
  This polynomial is scaled to match the experimental data and is drawn as a solid curve.
  }
  \label{fig:BgdswTHERM}
\end{figure}

The THERMINATOR 2 simulation has been shown to reproduce the background features in a $\uppi$K analysis~\cite{Kisiel:2017}. 
Figure~\ref{fig:BgdswTHERM} shows the THERMINATOR 2 simulation together with experimental data.  
The figure also shows a sixth-order polynomial fit to the simulation, as well as the fit polynomial scaled to match the data.
Clearly, the THERMINATOR 2 simulation provides a good description of the non-femtoscopic backgrounds in the \LamK systems, and can be used in a quantitative fashion to help fit the data.
More specifically, the non-femtoscopic backgrounds are modeled by sixth-order polynomial fits to THERMINATOR 2 simulation,
\begin{equation}
F_{\scaleto{\mathrm{THERM\; Bgd}}{6pt}}(k^{*}) = a{k^{*}}^{6}+ b{k^{*}}^{5} + c{k^{*}}^{4} + d{k^{*}}^{3} + e{k^{*}}^{2} + fk^{*} + g,
\label{eqn:BgdPoly}
\end{equation}
where the linear term coefficient is fixed to zero ($f=0$), and one polynomial is fit for each centrality class and \LamK charge combination.

Before fitting the signal region of the experimental data, the coefficients of each polynomial are fixed by fits to the THERMINATOR 2 background, shown in Fig.~\ref{fig:BgdswTHERM}.
The extracted polynomial is adjusted to best describe the experimental data by introducing a scale factor and a vertical shift,
\begin{equation}
F_{\scaleto{\mathrm{Bgd}}{6pt}}(k^{*}) = \alpha\times F_{\scaleto{\mathrm{THERM\; Bgd}}{6pt}}(k^{*}) + \beta,
\label{eqn:BgdScaleAndShift}
\end{equation}
where $\alpha$ and $\beta$ are determined by fitting to the data in the region $0.32 < k^{*} < 0.80$ GeV/$c$; all of the background parameters in Eq.~(\ref{eqn:BgdPoly}) and Eq.~(\ref{eqn:BgdScaleAndShift}) are fixed before fitting the low-\kstar signal region of the experimental correlation functions.
In all cases, the non-femtoscopic background correction was applied as a multiplicative factor to the correlation function during the fitting process.

An alternative approach for treating the non-femtoscopic background is to instead attempt to eliminate it.
The background can be effectively reduced by forming the reference distribution ($B(k^{*})$) with the ``Stavinskiy method"~\cite{Stavinskiy04, PhysRevD.82.052001} (see Appendix~\ref{App:StavMethod} for details).
With this method, mixed-event pairs are not used for the reference distribution; instead, same-event pseudo-pairs, formed by rotating one particle in a real pair by 180$^\circ$ in the transverse plane, are used.  
This rotation rids the pairs of any femtoscopic correlation, while maintaining correlations due to elliptic flow (and other suitably symmetric contributors).
The flattening effect of the method on the \LamKchP correlation functions can be seen in the Appendix~\ref{App:StavMethod}.

%************************************************************************************************************************
\subsection{Summarized correlation function construction}
\label{SummarizedFitProcedure}

The parameters included in the generation of a model correlation function are: $\lambda_{\mathrm{Fit}}$, $R$, $f_{0}$ ($\Re f_{0}$ and $\Im f_{0}$ separately), $d_{0}$, and normalization $\mathcal{N}$.
For the fit, a given pair and its conjugate (e.g., \LamKchP and \ALamKchM) share scattering parameters ($\Re f_{0}$, $\Im f_{0}$, $d_{0}$), and the three distinct analyses (\LamKchP, \LamKchM, and \LamKs) are assumed to have unique scattering parameters which are allowed to differ from each other.
The pair emission source for a given centrality class is assumed similar among all analyses; therefore, for each centrality, all \LamK analyses share a common radius parameter, $R$.
Furthermore, for each centrality class, a single $\lambda_{\mathrm{Fit}}$ parameter (see Eq.~(\ref{eqn:CfwRes2})) is shared amongst all.
Each fit correlation function has a unique normalization parameter, $\mathcal{N}$.

The experimental correlation functions were constructed separately for the two different field polarities applied by the ALICE L3 solenoid magnet during the data acquisition.
These are kept separate during the fitting process, and are combined using a weighted average when plotting, where the weight is the number of numerator pairs in the normalization range.
All experimental correlation functions are normalized in the range 0.32 $< k^{*} <$ 0.40 GeV/$c$, and fit in the range 0.0 $< k^{*} <$ 0.30 GeV/$c$.
For the \LamKchM analysis, the region 0.19 $< k^{*} <$ 0.23 GeV/$c$ was excluded from the fit to exclude the bump caused by the $\Omega^{-}$ decay.
A log-likelihood fit function is used as the statistic quantifying the quality of the fit to the experimental data~\cite{Lisa:2005dd}.

The complete fit function is constructed as follows.
The uncorrected, primary, fit correlation function, $C_{\Lambda\mathrm{K}}(k^{*}_{\mathrm{\Lambda K,\,True}})$, is constructed using Eq.~(\ref{eqn:LednickyEqn}).
Contributions from three parent systems which contribute via residual correlations are accounted for, as discussed in Sec.~\ref{ResidualCorrelations}.
The model correlation functions describing these parent systems, $C_{ij}(k^{*}_{ij,\,\mathrm{True}})$, are obtained using Eq.~(\ref{eqn:LednickyEqn}) for Coulomb-neutral pairs or experimental data for \XiKpm contributions.
The residual contributions are then found by running each parent correlation function through the appropriate transform matrix, via Eq.~(\ref{eqn:ResidualsTransform}).
The model primary and residual contributions are combined, via Eq.~(\ref{eqn:CfwRes}) with the $\lambda_{ij}$ values listed in Tab.~\ref{tab:LambdaValues_3Res}, to form $C'_{\mathrm{Fit}}$(\ktrue).
Corrections are applied to account for finite track momentum resolution effects using Eq.~(\ref{eqn:MomResCorrection}), to obtain $C'_{\mathrm{Fit}}(k^{*}_{\mathrm{Rec}})$.
Finally, the non-femtoscopic background correction, $F_{\mathrm{Bgd}}(k^{*}_{\mathrm{Rec}})$, is applied as described in Sec.~\ref{NonFlatBackground}, and the final fit function is obtained,
\begin{equation}
C_{\mathrm{Fit}}(k^{*}_{\mathrm{Rec}}) = \mathcal{N}\times F_{\mathrm{Bgd}}(k^{*}_{\mathrm{Rec}})\times C'_{\mathrm{Fit}}(k^{*}_{\mathrm{Rec}}),
\end{equation}
where $\mathcal{N}$ is a normalization parameter.
This model correlation function is then fit to the experimental correlation function.

%************************************************************************************************************************
\subsection{Systematic uncertainties}
\label{SysErrs}

To estimate the systematic uncertainties in the analysis, the selection criteria were varied, and experimental correlation functions and fit results were obtained for each variation.
To quantify the systematic uncertainties on the data points, the experimental correlation functions from each variation of the selection criteria were combined, giving a distribution of values for each \kstar.
From these distributions, the standard deviations were calculated and assigned as the systematic uncertainties of the corresponding data points.

A similar process was followed for estimating the systematic uncertainties of the extracted fit parameters.
Namely, the extracted fit parameters from each variation were averaged, and the resulting standard deviations taken as the systematic uncertainties.
Additionally, a systematic analysis was done on the fit method through varying the \kstar fit range, varying the modeling of the non-femtoscopic background, as well as varying $\tau_{\mathrm{max}}$ in the treatment of residual correlations.
The choice of \kstar fit range was varied by $\pm$ 25\%.
In addition to modeling with a polynomial fit to the THERMINATOR 2 simulation, the backgrounds of all of the systems were modeled by fitting to the data with a linear, quadratic, and Gaussian form.
Finally, $\tau_{\mathrm{max}}$ was varied from the default value of $\tau_{\mathrm{max}} = 10$ fm/$c$ down to $\tau_{\mathrm{max}} = 6$ fm/$c$ and up to $\tau_{\mathrm{max}} = 15$ fm/$c$.
The resulting uncertainties in the extracted parameter sets were combined with the uncertainties arising from the variations of the selection criteria.
The systematic uncertainties of the extracted parameters sets are due primarily to the fit method variations, i.e., the selection criteria do not contribute significantly.

%************************************************************************************************************************
%************************************************************************************************************************
%\clearpage
\section{Results}
\label{sec:Results}

\begin{figure}[h!]
  \centering
  \includegraphics[width=\linewidth]{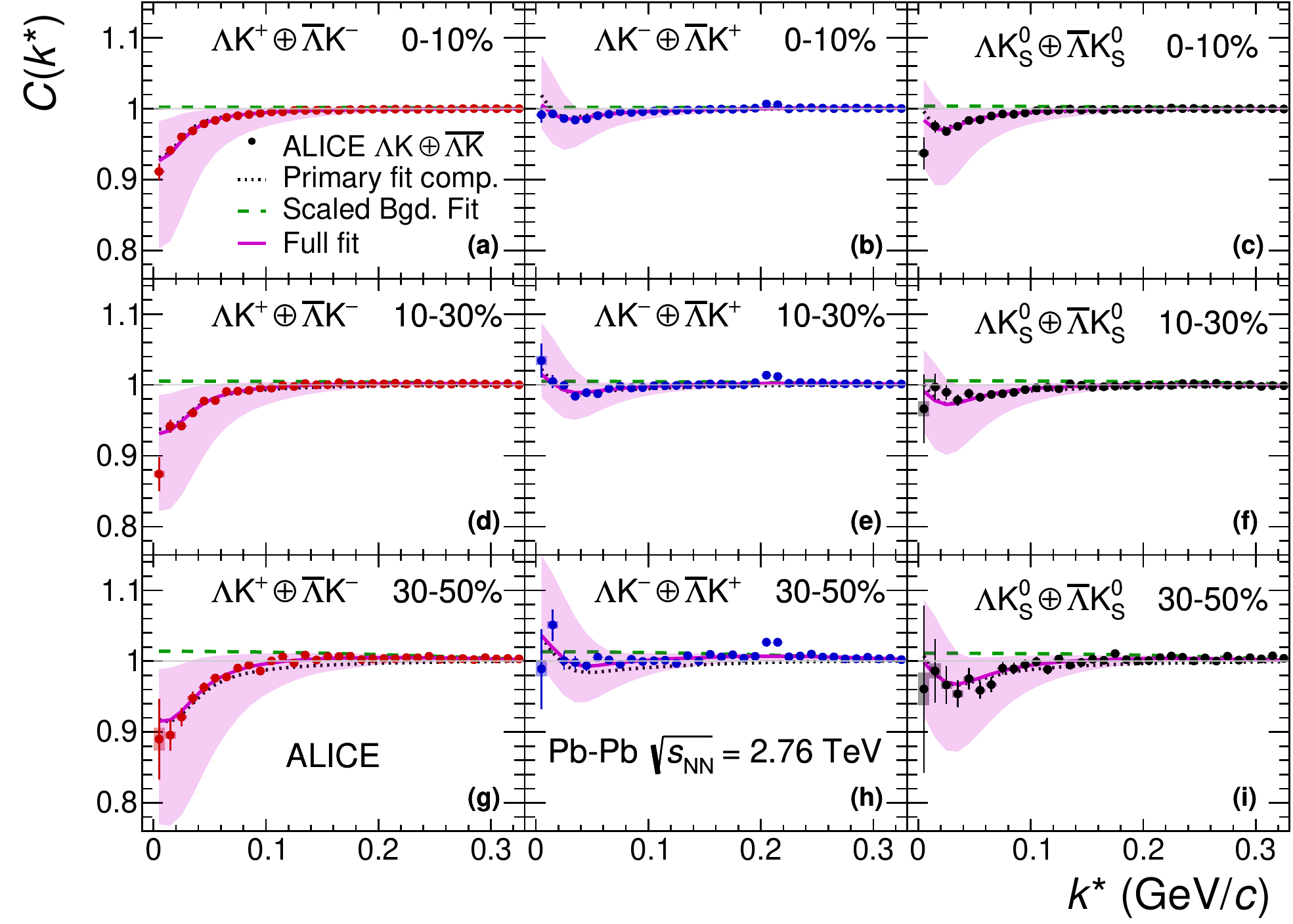}
  \caption[\LamK data with fits]
  {
  (Color online) Fit results for the \LamK data, with pair and conjugate combined.
  The \LamKchP$\oplus$\ALamKchM data are shown in the left column, the \LamKchM$\oplus$\ALamKchP in the middle, and the \LamKs$\oplus$\ALamKs in the right. 
  Rows differentiate the different centrality intervals (0--10\% in the top, 10--30\% in the middle, and 30--50\% in the bottom).
  Lines represent statistical uncertainties, while boxes represent systematic uncertainties.
  The dotted curve shows the primary (\LamK) contribution to the fit, the dashed curve shows the fit to the non-femtoscopic background, and the solid curve shows the final fit.
  The shaded region represents the (combined statistical and systematic) uncertainty of the final fit.
 }
  \label{fig:LamKFits_3Res}
\end{figure}

Figure~\ref{fig:LamKFits_3Res} shows the \LamK data with fits for all studied centrality percentile intervals (0--10\%, 10--30\%, and 30--50\%). 
All six \LamK systems (\LamKchP, \ALamKchM, \LamKchM, \ALamKchP, \LamKs, \ALamKs) are fit simultaneously across all centralities, with a single radius and normalization $\lambda_{\mathrm{Fit}}$ parameter for each centrality interval. 
The figure shows the primary (\LamK) contribution to the fit (i.e., $1 + \lambda'_{\Lambda\mathrm{K}}C_{\Lambda\mathrm{K}}(k^{*}_{\Lambda\mathrm{K}})$ in Eq.~(\ref{eqn:CfwRes})), the fit to the non-femtoscopic background, and the final fit, with all residual contributions included and after all corrections have been applied.
The extraction of the primary \LamK component is the purpose of this study.
The figure demonstrates that the final fit function is similar to the primary \LamK component, with the largest differences between the two observed in the 30--50\% centrality interval due mainly to the large contribution of the non-femtoscopic background.

\begin{figure}[htp]
  \centering
  %%----start of first subfigure---
  \subfigure{
    \label{fig:ScattParams_3Res:a}
    \includegraphics[width=0.49\linewidth]{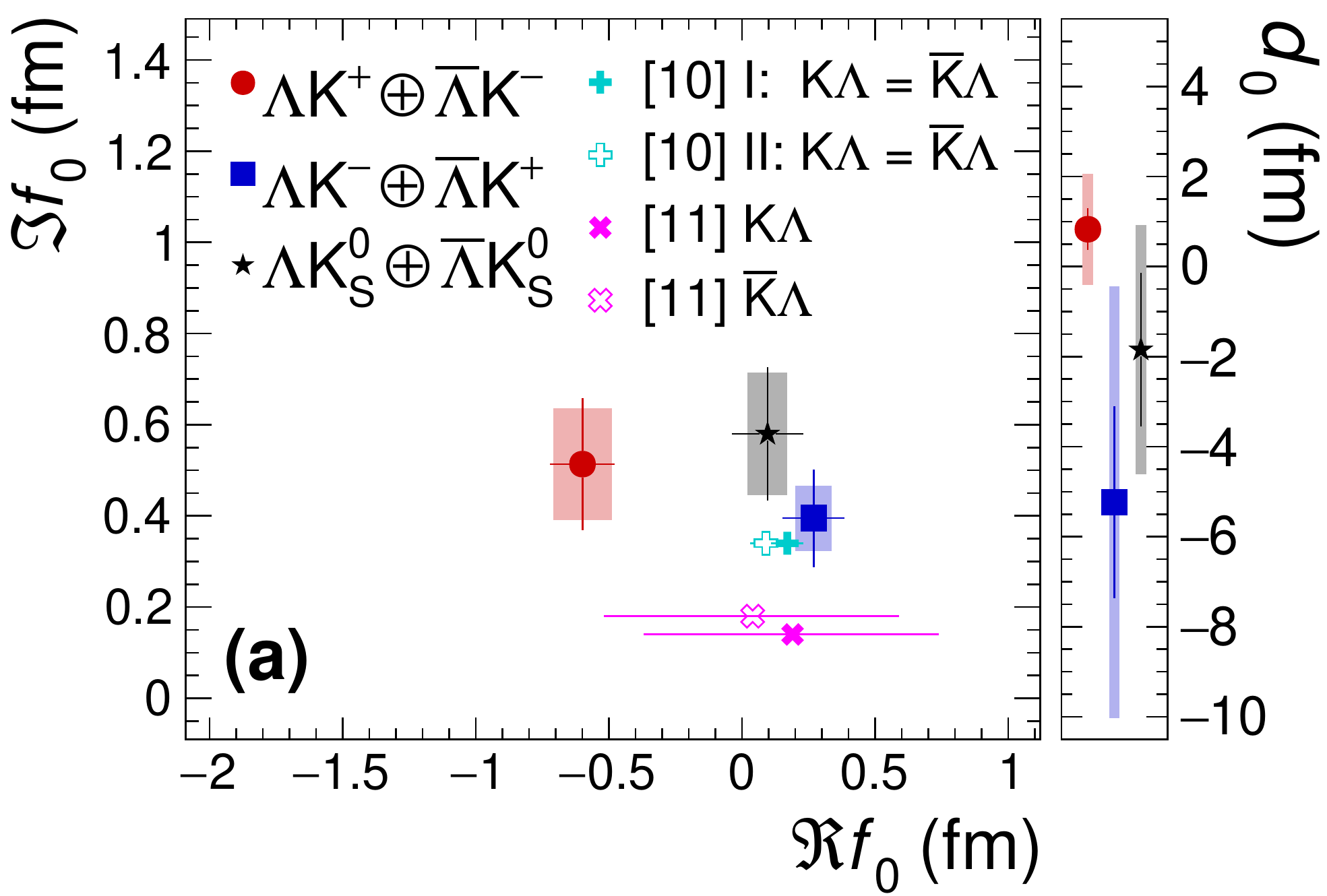}}
  %%----start of second subfigure---  
  \subfigure{
    \label{fig:ScattParams_3Res:b}
    \includegraphics[width=0.49\linewidth]{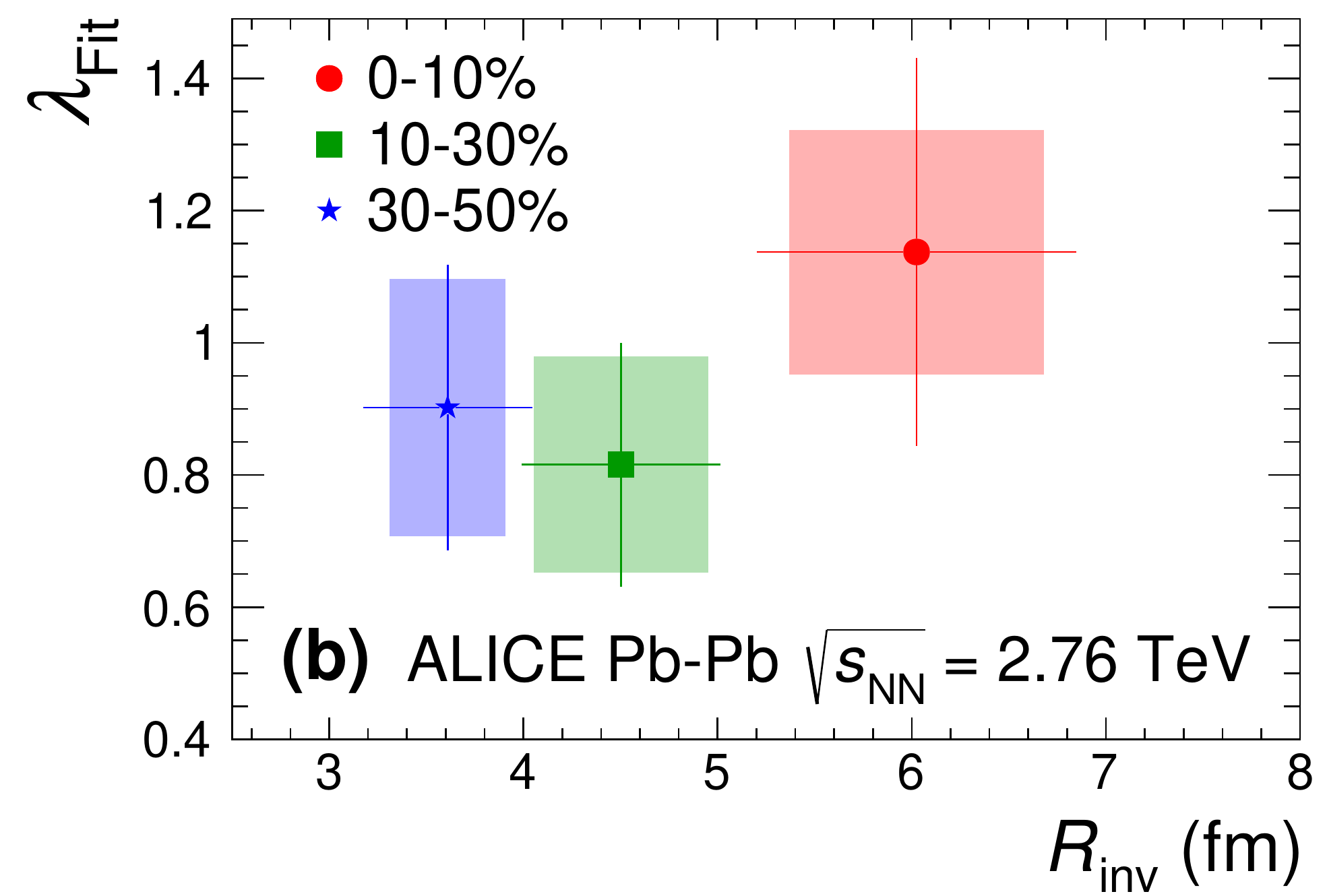}}
  %%----overall caption----
  \caption[Extracted Scattering Parameters]
  {
  (Color online) Extracted fit parameters for all of the \LamK systems.
  In the figures, lines represent statistical uncertainties, while boxes represent systematic uncertainties.  
  (Left) The scattering parameters, $\Im f_{0}$ and $\Re f_{0}$, together with $d_{0}$ to the right, for the \LamKchP (circles), \LamKchM (squares) and \LamKs (stars) systems.  
  (Right) The $\lambda_{\mathrm{Fit}}$ and radius parameters for the 0--10\% (circles), 10--30\% (squares), and 30--50\% (stars) centrality intervals.  
  In the fit, all \LamK systems share common radii.
  The cross~\cite{Liu:2006xja} and X~\cite{Mai:2009ce} points show theoretical predictions made using chiral perturbation theory.
  }  
  \label{fig:ScattParams_3Res}
\end{figure}

Figure~\ref{fig:ScattParams_3Res} (left) summarizes the extracted \LamK scattering parameters, and includes theoretical predictions made using chiral perturbation theory~\cite{Liu:2006xja,Mai:2009ce}.
For all \LamK systems, positive imaginary parts of the scattering lengths, $\Im(f_{0})$, are extracted from the experimental data. 
This is expected, as $\Im(f_{0})$ describes the inelastic scattering channels.
More interestingly, the results show that the \LamKchP and \LamKchM systems differ in the sign of the real part, $\Re(f_{0})$, of their scattering lengths, with a negative value for \LamKchP and positive value for \LamKchM.
The $\Re f_{0}$ extracted for the \LamKs system is positive, and within uncertainties of that of \LamKchM. 
The real part of the scattering length describes the effect of the strong interaction: a positive $\Re(f_{0})$ signifies that the interaction is attractive, while a negative $\Re(f_{0})$ signifies a repulsive interaction, as is the usual convention in femtoscopy.
Therefore, the femtoscopic signals from this analysis demonstrate that the strong interaction acts repulsively in the \LamKchP system and attractively in the \LamKchM system.
The analysis suggests that the \LamKs interaction is attractive, however the uncertainty of the result does not permit a definite conclusion.
Finally, the results indicate that the effective range of the interaction, $d_{0}$, is positive in the \LamKchP system and negative in the \LamKchM and \LamKs systems.

In Fig.~\ref{fig:ScattParams_3Res} (left), the predictions of~\cite{Liu:2006xja} do not distinguish the K\Lam and K\ALam interactions and results are shown for two different parameter sets, whereas~\cite{Mai:2009ce} offers unique K\Lam and $\overline{\mathrm{K}}$\Lam scattering parameters for a single parameter set. 
Past studies of kaon-proton scattering found the K$^{-}$p interaction to be attractive, and that of the K$^{+}$p to be repulsive~\cite{Humphrey:1962zz, Hadjimichef:2002xe, Ikeda:2012au, PhysRevLett.124.092301}.
With respect to the kaons, this is similar to the current finding of an attractive \LamKchM interaction and a repulsive \LamKchP interaction.
This difference could arise from different quark--antiquark interactions between the pairs ($\rm s\overline{s}$ in \LamKchP, $\rm u\overline{u}$ in \LamKchM).
A related explanation could be that the effect is due to the different net strangeness for each system.
The quark content of the \Lam (\ALam) is uds ($\overline{\mathrm{uds}}$), that of the \KchP (\KchM) is u$\overline{\mathrm{s}}$ ($\overline{\mathrm{u}}$s), and the \Ks is a mixture of the neutral $\mathrm{K}^{0}$ and $\overline{\mathrm{K}^{0}}$ states with quark content $\frac{1}{\sqrt{2}}\left[\mathrm{d\overline{s} + \overline{d}s}\right]$.
It is interesting to note the presence of a $\mathrm{s\overline{s}}$ pair in the \LamKchP system contrasted with a $\mathrm{u\overline{u}}$ pair in the \LamKchM system.
Additionally, although the \Ks is an average of \KchP and \KchM in some respects (e.g., electrically), it contains (anti)down quarks, whereas the \Kpm contain (anti)up quarks.

\begin{figure}[h]
  \centering
  \includegraphics[width=0.60\textwidth]{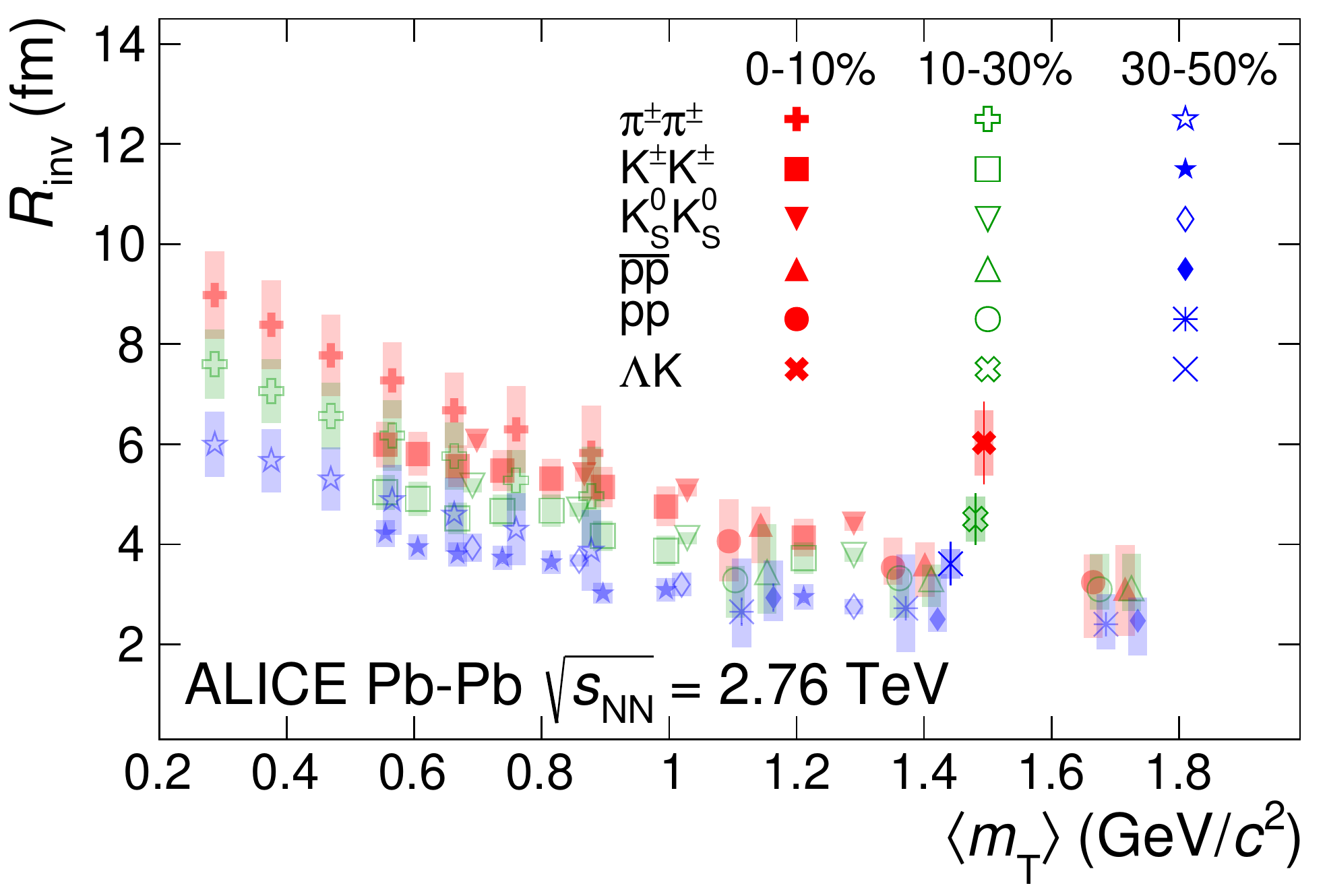}
  \caption[\mt Scaling of Radii: 3 Residuals in Fit]
  {
  (Color online) Extracted fit $R_{\mathrm{inv}}$ parameters as a function of pair transverse mass (\mt) for several centralities.
  Results from the \LamK analysis are presented together with ALICE published data~\cite{Adam:2015vja} for various other pair systems. 
  Statistical (lines) and systematic (boxes) uncertainties are shown. 
  }
  \label{fig:mTScalingOfRadii_3Res}
\end{figure}

Figure~\ref{fig:ScattParams_3Res} (right) presents the $\lambda_{\mathrm{Fit}}$ and radius parameters for all three studied centrality percentile ranges.
The $\lambda_{\mathrm{Fit}}$ parameters are expected to be close to unity. 
A comparison of the extracted radii from this study to those of other systems measured by ALICE~\cite{Adam:2015vja} is shown in Fig.~\ref{fig:mTScalingOfRadii_3Res}. 
The figure shows $R_{\mathrm{inv}}$ as a function of \mt for several centrality ranges and for several different pair systems.
The \mt value used for the present \LamK results was taken as the average of the three systems.
For non-identical particle pairs, to be more directly analogous to the single particle \mt, the definition of the pair transverse mass used in this study is
\begin{equation}
\begin{aligned}
 m_{\mathrm{T, pair}}^{2} &= \left( \frac{m_{\mathrm{inv}}}{2} \right)^{2} + \left( \frac{1}{2} |\textbf{\textit{p}}_{\mathrm{T,1}} + \textbf{\textit{p}}_{\mathrm{T,2}}| \right)^{2} = (K^{0})^{2} - (K^{3})^{2}, \quad \mathrm{where} ~~K^{\mu} \equiv \frac{1}{2} \left( p_{1}^{\mu} + p_{2}^{\mu} \right).
\end{aligned}
\label{eqn:PairmTv1}
\end{equation}
The radii are observed to increase for more central events, as expected from a simple geometric picture of the collisions.
Femtoscopy probes the distribution of relative positions of outgoing particles whose velocities have a specific magnitude and direction~\cite{Lisa:2005dd}, referred to as ``regions of homogeneity''~\cite{Akkelin:1995gh}.
Consequently, for each pair system, the radii decrease with increasing \mt, as expected in the presence of collective radial flow~\cite{Akkelin:1995gh}.
It was found that~\cite{Kisiel:2014upa}, even in the presence of global \mt-scaling for the three-dimensional radii in the Longitudinally Co-Moving System (LCMS), a particle species dependence will exist for the $R_{\mathrm{inv}}$ measured in the PRF, due to trivial kinematic reasons.
These kinematic effects, resulting from the transformation from LCMS to PRF, cause smaller masses to exhibit larger $R_{\mathrm{inv}}$~\cite{Adam:2015vja} (explaining, for instance, why the pion radii are systematically higher than kaon radii at the same approximate \mt).

It is clear from the results in Fig.~\ref{fig:mTScalingOfRadii_3Res} that the \LamK systems do not conform to the approximate \mt-scaling of the identical particle pair source sizes.
There are two important consequences of the hydrodynamic nature of the system to consider when interpreting non-identical femtoscopic results.
First, the hydrodynamic response of the medium produces the approximate \mt-scaling with respect to the single-particle sources.
Second, this response confines higher-\mt particles to smaller homogeneity regions and pushes their average emission points further in the ``out" direction~\cite{Retiere:2003kf} in a coordinate system chosen according to the out-side-long prescription (where the ``long" axis is parallel to the beam, ``out" is parallel to the total transverse momentum of the pair, and ``side" is orthogonal to both).
For identical particle studies, in which the pair source is comprised of two identical single particle sources commonly affected by the space-time shift, the femtoscopic radii naturally follow the \mt-scaling trend.
However, for the case of non-identical particles, the pair emission source is a superposition of two unique single-particle sources, which are affected differently by the hydrodynamic response of the system.
Therefore, the \Lam and K sources differ both in size and space--time location, leading to an ``emission asymmetry", with the \Lam source both smaller in size and further out in the fireball than that of the kaons.

\begin{figure}[htp]
  \centering
  %%----start of first subfigure---
  \subfigure{
    \label{fig:LamKchP_ReC00C11_0010:a}
    \includegraphics[width=0.49\linewidth]{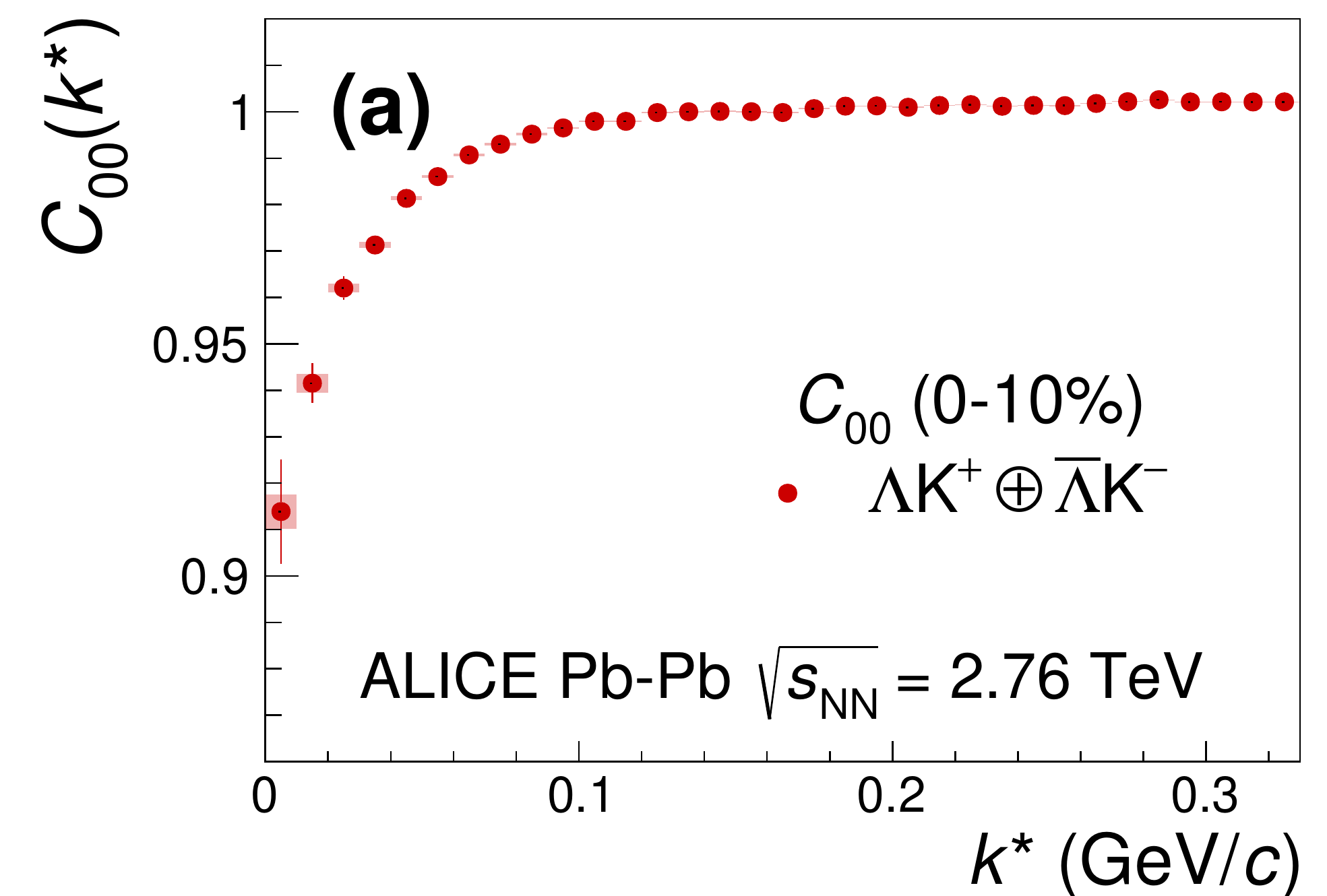}}
  %%----start of second subfigure---  
  \subfigure{
    \label{fig:LamKchP_ReC00C11_0010:b}
    \includegraphics[width=0.49\linewidth]{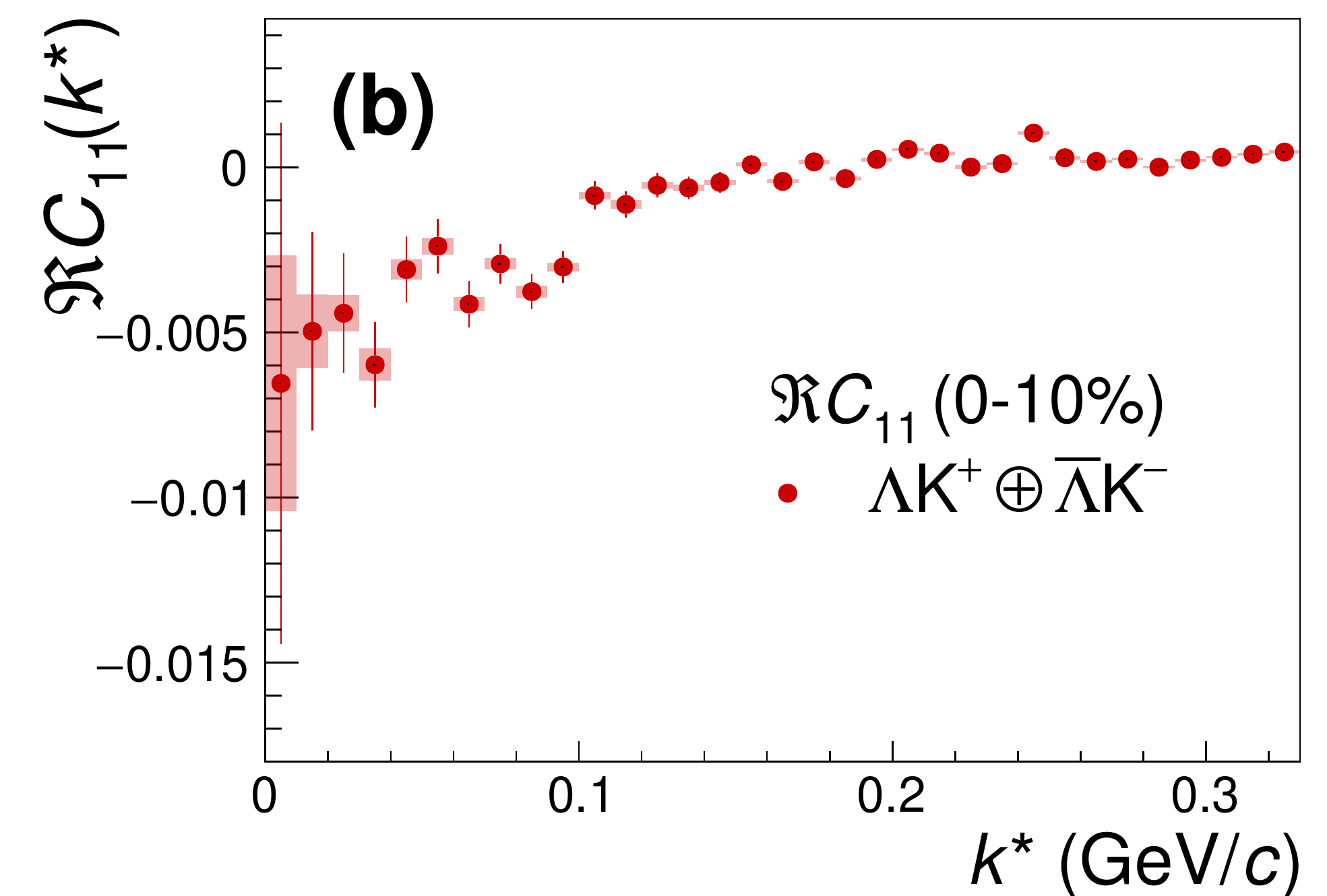}}
  %%----overall caption----
  \caption[\LamKchP $C_{00}$ and $\Re C_{11}$ Spherical Harmonic Components (0--10\%)]
  {
  (Color online) Spherical harmonics components $C_{00}$ (left) and $\Re C_{11}$ (right) of the \LamKchP correlation function for the 0--10\% centrality interval.  
  Statistical (lines) and systematic (boxes) uncertainties are shown.
The $C_{00}$ component is similar to the one-dimensional correlation functions typically studied, and probes the overall size of the source.
The $\Re C_{11}$ component probes the asymmetry in the system; a non-zero value reveals the asymmetry.
  }  
  \label{fig:LamKchP_ReC00C11_0010}
\end{figure}

A separation of the single-particle sources in the ``out" direction is expected for \LamK pairs at midrapidity in Pb--Pb collisions, as described above, and the experimental data support such an emission asymmetry.
In addition to the ``size" of the emitting region (more precisely, the second moments of the emission functions) accessible with identical particle studies, non-identical particle correlations are sensitive to the relative emission shifts, i.e., the first moments of the emission function~\cite{Kisiel:2009eh}.
The spherical harmonic decomposition of the correlation function offers an elegant method for extracting information about the emission asymmetries~\cite{Chajecki:2008vg, PhysRevC.72.054902, Kisiel:2009iw}.
With this method, one can draw a wealth of information from just a few components of the decomposition.
Particularly, the $l=0$, $m=0$ component, $C_{00}$, quantifies the angle-integrated strength of the correlation function, and probes the overall size of the source.
Of interest here, the real part of the $l=1$, $m=1$ component, $\Re C_{11}$, probes the asymmetry of the system in the ``out" direction; a non-zero value reveals the asymmetry. 
Figure~\ref{fig:LamKchP_ReC00C11_0010} shows the $C_{00}$ and $\Re C_{11}$ components from the spherical decomposition of the \LamKchP data in the 0--10\% centrality interval.
The $\Re C_{11}$ component shows a clear deviation from zero, and the negative value signifies that the \Lam particles are, on average, emitted further out and/or earlier than the K mesons.
This conclusion is supported by the results obtained from the THERMINATOR 2 model, shown in Fig.~\ref{fig:LamKchP_StdThermSources}.
Furthermore, this emission asymmetry effect can inflate the radii extracted with the one-dimensional Lednick\'y model, which assumes a spherically symmetric source with no offsets (i.e., $R_{\mathrm{out}} = R_{\mathrm{side}} = R_{\mathrm{long}}$ and $\mu_{\mathrm{out}} = \mu_{\mathrm{side}} = \mu_{\mathrm{long}} = 0$).
This effect is demonstrated in Appendix~\ref{App:THERM} using the THERMINATOR 2 simulation.
In Fig.~\ref{fig:mTScalingOfRadii_3Res}, the largest violation of the \mt-scaling for the \LamK system is observed for the 0--10\% centrality interval, in which one expects the largest emission asymmetry.

%************************************************************************************************************************
%************************************************************************************************************************
\section{Summary}
\label{sec:Summary}

\begin{table}[htbp]
 \centering
 \caption{
 Extracted fit parameters.
 The uncertainties marked as ``stat." are those returned by MINUIT~\cite{JAMES1975343}, and those marked as ``syst." result from the systematic analysis.
 }
 \resizebox{\columnwidth}{!}{
 \renewcommand{\arraystretch}{1.3}
 \begin{tabular}{c|c|c|c}
  \clineB{1-3}{3.0}
  Centrality & $\lambda_{\mathrm{Fit}}$ & \multicolumn{1}{c}{$R_{\mathrm{inv}}$} & \multicolumn{1}{c}{} \\
  \clineB{1-3}{3.0}  
   0--10\%
     & \ArrLamKchP(1) $\pm$ \ArrLamKchP(2) (stat.) $\pm$ \ArrLamKchP(3) (syst.)    %Lambda (LamKchP 0010)
     & \multicolumn{1}{c}{\ArrLamKchP(4) $\pm$ \ArrLamKchP(5) (stat.) $\pm$ \ArrLamKchP(6) (syst.)}    %Radius (LamKchP & ALamKchM 0010)
     & \multicolumn{1}{c}{} \\    
   \cline{1-3}
   
   10--30\%
     & \ArrLamKchP(7) $\pm$ \ArrLamKchP(8) (stat.) $\pm$ \ArrLamKchP(9) (syst.)      %Lambda (LamKchP 1030)
     & \multicolumn{1}{c}{\ArrLamKchP(10) $\pm$ \ArrLamKchP(11) (stat.) $\pm$ \ArrLamKchP(12) (syst.)}    %Radius (LamKchP & ALamKchM 1030)    
     & \multicolumn{1}{c}{} \\  
   \cline{1-3}           
   
   30--50\%
     & \ArrLamKchP(13) $\pm$ \ArrLamKchP(14) (stat.) $\pm$ \ArrLamKchP(15) (syst.)    %Lambda (LamKchP 3050)
     & \multicolumn{1}{c}{\ArrLamKchP(16) $\pm$ \ArrLamKchP(17) (stat.) $\pm$ \ArrLamKchP(18) (syst.)}    %Radius (LamKchP & ALamKchM 3050)   
     & \multicolumn{1}{c}{} \\  
   %\clineB{1-3}{6.0}
   \cline{1-3}
  %%%%%%%%%%%%%%%%%%%%%%%%%%%%%%%%%%%%%%%%%%%%%%%%%%%%%%%%%%%%%%%%%%%%%%%%%%%%%%%%%%%%%%%%%%%%%%%%%%%%%%%%%%%%%%%%%%%%%%%%%%%%%%%%% 
   \multicolumn{1}{c}{} & \multicolumn{1}{c}{} & \multicolumn{1}{c}{} & \multicolumn{1}{c}{} \\
   %\hlineB{6.0}
   \hlineB{3.0}
  System & $\Re f_{0}$ & $\Im f_{0}$ & $d_{0}$ \\
  \clineB{1-4}{3.0}  
   \LamKchP $\oplus$ \ALamKchM 
     & \ArrLamKchP(19) $\pm$ \ArrLamKchP(20) (stat.) $\pm$ \ArrLamKchP(21) (syst.)   %Ref0   (LamKchP & ALamKchM)
     & \ArrLamKchP(22) $\pm$ \ArrLamKchP(23) (stat.) $\pm$ \ArrLamKchP(24) (syst.)    %Imf0   (LamKchP & ALamKchM)
     & \ArrLamKchP(25) $\pm$ \ArrLamKchP(26) (stat.) $\pm$ \ArrLamKchP(27) (syst.) \\ %d0     (LamKchP & ALamKchM)
     
   \hline
   
   \LamKchM $\oplus$ \ALamKchP
     & \ArrLamKchM(19) $\pm$ \ArrLamKchM(20) (stat.) $\pm$ \ArrLamKchM(21) (syst.)   %Ref0   (LamKchM & ALamKchP)
     & \ArrLamKchM(22) $\pm$ \ArrLamKchM(23) (stat.) $\pm$ \ArrLamKchM(24) (syst.)    %Imf0   (LamKchM & ALamKchP)
     & \ArrLamKchM(25) $\pm$ \ArrLamKchM(26) (stat.) $\pm$ \ArrLamKchM(27) (syst.) \\ %d0     (LamKchM & ALamKchP)     
   \hline           
   
   \LamKs $\oplus$ \ALamKs
     & \ArrLamKs(19) $\pm$ \ArrLamKs(20) (stat.) $\pm$ \ArrLamKs(21) (syst.)   %Ref0   (LamKchM & ALamKchP)
     & \ArrLamKs(22) $\pm$ \ArrLamKs(23) (stat.) $\pm$ \ArrLamKs(24) (syst.)    %Imf0   (LamKchM & ALamKchP)
     & \ArrLamKs(25) $\pm$ \ArrLamKs(26) (stat.) $\pm$ \ArrLamKs(27) (syst.) \\ %d0     (LamKchM & ALamKchP)     
   \hline   
     
 \end{tabular}
 }
 %\caption{
 %Extracted fit parameters.
 %The uncertainties marked as ``stat." are those returned by MINUIT, and those marked as ``syst." result from the systematic analysis.
 %}
 \label{tab:FitResultsLamK_3Res}
\end{table}

Results from a femtoscopic analysis of \LamK correlations in Pb--Pb collisions at $\sqrt{s_{\mathrm{NN}}}$ = 2.76 TeV measured by the ALICE experiment at the LHC have been presented, and are summarized in Table~\ref{tab:FitResultsLamK_3Res}.
The femtoscopic radii, $\lambda$ parameters, and scattering parameters were extracted from one-dimensional correlation functions in terms of the invariant momentum difference.
The scattering parameters of \LamK pairs in all three charge combinations (\LamKchP, \LamKchM, and \LamKs) were measured for the first time.
The non-femtoscopic backgrounds observed in the experimental data were described quantitatively with the THERMINATOR 2 event generator, and were found to result almost entirely from collective effects.
Striking differences are observed in the \LamKchP, \LamKchM, and \LamKs correlation functions, which are reflected in the unique set of scattering parameters extracted for each.
These scattering parameters indicate that the strong force is repulsive in the \LamKchP interaction and attractive in the \LamKchM interaction, and suggest that the interaction is also attractive in the \LamKs system.
This effect could be due to different quark--antiquark interactions between the pairs, or from different net strangeness present in each system. 
The extracted source radii describing the \LamK systems are larger than expected from naive extrapolation of identical particle femtoscopic studies.
This effect is interpreted as resulting from the separation in space--time of the single-particle \Lam and K source distributions (i.e., the emission asymmetry of the source), which is confirmed by the spherical harmonics decomposition of the correlation functions.
%\clearpage

%%%%% acknowledgements
\newenvironment{acknowledgement}{\relax}{\relax}
\begin{acknowledgement}
\section*{Acknowledgements}
% Version: 2020-04-29

The ALICE Collaboration would like to thank all its engineers and technicians for their invaluable contributions to the construction of the experiment and the CERN accelerator teams for the outstanding performance of the LHC complex.
The ALICE Collaboration gratefully acknowledges the resources and support provided by all Grid centres and the Worldwide LHC Computing Grid (WLCG) collaboration.
The ALICE Collaboration acknowledges the following funding agencies for their support in building and running the ALICE detector:
A. I. Alikhanyan National Science Laboratory (Yerevan Physics Institute) Foundation (ANSL), State Committee of Science and World Federation of Scientists (WFS), Armenia;
Austrian Academy of Sciences, Austrian Science Fund (FWF): [M 2467-N36] and Nationalstiftung f\"{u}r Forschung, Technologie und Entwicklung, Austria;
Ministry of Communications and High Technologies, National Nuclear Research Center, Azerbaijan;
Conselho Nacional de Desenvolvimento Cient\'{\i}fico e Tecnol\'{o}gico (CNPq), Financiadora de Estudos e Projetos (Finep), Funda\c{c}\~{a}o de Amparo \`{a} Pesquisa do Estado de S\~{a}o Paulo (FAPESP) and Universidade Federal do Rio Grande do Sul (UFRGS), Brazil;
Ministry of Education of China (MOEC) , Ministry of Science \& Technology of China (MSTC) and National Natural Science Foundation of China (NSFC), China;
Ministry of Science and Education and Croatian Science Foundation, Croatia;
Centro de Aplicaciones Tecnol\'{o}gicas y Desarrollo Nuclear (CEADEN), Cubaenerg\'{\i}a, Cuba;
Ministry of Education, Youth and Sports of the Czech Republic, Czech Republic;
The Danish Council for Independent Research | Natural Sciences, the VILLUM FONDEN and Danish National Research Foundation (DNRF), Denmark;
Helsinki Institute of Physics (HIP), Finland;
Commissariat \`{a} l'Energie Atomique (CEA) and Institut National de Physique Nucl\'{e}aire et de Physique des Particules (IN2P3) and Centre National de la Recherche Scientifique (CNRS), France;
Bundesministerium f\"{u}r Bildung und Forschung (BMBF) and GSI Helmholtzzentrum f\"{u}r Schwerionenforschung GmbH, Germany;
General Secretariat for Research and Technology, Ministry of Education, Research and Religions, Greece;
National Research, Development and Innovation Office, Hungary;
Department of Atomic Energy Government of India (DAE), Department of Science and Technology, Government of India (DST), University Grants Commission, Government of India (UGC) and Council of Scientific and Industrial Research (CSIR), India;
Indonesian Institute of Science, Indonesia;
Centro Fermi - Museo Storico della Fisica e Centro Studi e Ricerche Enrico Fermi and Istituto Nazionale di Fisica Nucleare (INFN), Italy;
Institute for Innovative Science and Technology , Nagasaki Institute of Applied Science (IIST), Japanese Ministry of Education, Culture, Sports, Science and Technology (MEXT) and Japan Society for the Promotion of Science (JSPS) KAKENHI, Japan;
Consejo Nacional de Ciencia (CONACYT) y Tecnolog\'{i}a, through Fondo de Cooperaci\'{o}n Internacional en Ciencia y Tecnolog\'{i}a (FONCICYT) and Direcci\'{o}n General de Asuntos del Personal Academico (DGAPA), Mexico;
Nederlandse Organisatie voor Wetenschappelijk Onderzoek (NWO), Netherlands;
The Research Council of Norway, Norway;
Commission on Science and Technology for Sustainable Development in the South (COMSATS), Pakistan;
Pontificia Universidad Cat\'{o}lica del Per\'{u}, Peru;
Ministry of Science and Higher Education, National Science Centre and WUT ID-UB, Poland;
Korea Institute of Science and Technology Information and National Research Foundation of Korea (NRF), Republic of Korea;
Ministry of Education and Scientific Research, Institute of Atomic Physics and Ministry of Research and Innovation and Institute of Atomic Physics, Romania;
Joint Institute for Nuclear Research (JINR), Ministry of Education and Science of the Russian Federation, National Research Centre Kurchatov Institute, Russian Science Foundation and Russian Foundation for Basic Research, Russia;
Ministry of Education, Science, Research and Sport of the Slovak Republic, Slovakia;
National Research Foundation of South Africa, South Africa;
Swedish Research Council (VR) and Knut \& Alice Wallenberg Foundation (KAW), Sweden;
European Organization for Nuclear Research, Switzerland;
Suranaree University of Technology (SUT), National Science and Technology Development Agency (NSDTA) and Office of the Higher Education Commission under NRU project of Thailand, Thailand;
Turkish Atomic Energy Agency (TAEK), Turkey;
National Academy of  Sciences of Ukraine, Ukraine;
Science and Technology Facilities Council (STFC), United Kingdom;
National Science Foundation of the United States of America (NSF) and United States Department of Energy, Office of Nuclear Physics (DOE NP), United States of America.    %%%%%%% done by webmaster team
\end{acknowledgement}

%%%%%%%% Bibliography (In case of using bibtex generate the bbl requested by arXiv)
\bibliographystyle{utphys}   % Remember we use title in the biblio
\bibliography{LamK_bibfile}

\clearpage

%************************************************************************************************************************
%************************************************************************************************************************
\section{The ALICE Collaboration}
\label{app:collab}
% Collaboration: CERN-LHC-ALICE
% Generation Date is 2020-04-14

% How to use:
%%%%%%%%% appendix with author list
%\appendix
%\section{The ALICE Collaboration}
%\label{app:collab}
%\input{Alice_Authorslist_XXXX-Axx-XX.tex}
\begingroup
\small
\begin{flushleft}
S.~Acharya\Irefn{org141}\And 
D.~Adamov\'{a}\Irefn{org95}\And 
A.~Adler\Irefn{org74}\And 
J.~Adolfsson\Irefn{org81}\And 
M.M.~Aggarwal\Irefn{org100}\And 
G.~Aglieri Rinella\Irefn{org34}\And 
M.~Agnello\Irefn{org30}\And 
N.~Agrawal\Irefn{org10}\textsuperscript{,}\Irefn{org54}\And 
Z.~Ahammed\Irefn{org141}\And 
S.~Ahmad\Irefn{org16}
% ...
\And
\renewcommand\labelenumi{\textsuperscript{\theenumi}~}

\section*{Affiliation notes}
\renewcommand\theenumi{\roman{enumi}}
\begin{Authlist}
\item \Adef{org*}Deceased
\item \Adef{orgI}Italian National Agency for New Technologies, Energy and Sustainable Economic Development (ENEA), Bologna, Italy
\item \Adef{orgII}Dipartimento DET del Politecnico di Torino, Turin, Italy
\item \Adef{orgIII}M.V. Lomonosov Moscow State University, D.V. Skobeltsyn Institute of Nuclear, Physics, Moscow, Russia
\item \Adef{orgIV}Department of Applied Physics, Aligarh Muslim University, Aligarh, India
\item \Adef{orgV}Institute of Theoretical Physics, University of Wroclaw, Poland
\end{Authlist}

\section*{Collaboration Institutes}
\renewcommand\theenumi{\arabic{enumi}~}
\begin{Authlist}
\item \Idef{org1}A.I. Alikhanyan National Science Laboratory (Yerevan Physics Institute) Foundation, Yerevan, Armenia
\item \Idef{org2}Bogolyubov Institute for Theoretical Physics, National Academy of Sciences of Ukraine, Kiev, Ukraine
\item \Idef{org3}Bose Institute, Department of Physics  and Centre for Astroparticle Physics and Space Science (CAPSS), Kolkata, India
\item \Idef{org4}Budker Institute for Nuclear Physics, Novosibirsk, Russia
\item \Idef{org5}California Polytechnic State University, San Luis Obispo, California, United States
\item \Idef{org6}Central China Normal University, Wuhan, China
\item \Idef{org7}Centre de Calcul de l'IN2P3, Villeurbanne, Lyon, France
\item \Idef{org8}Centro de Aplicaciones Tecnol\'{o}gicas y Desarrollo Nuclear (CEADEN), Havana, Cuba
\item \Idef{org9}Centro de Investigaci\'{o}n y de Estudios Avanzados (CINVESTAV), Mexico City and M\'{e}rida, Mexico
\item \Idef{org10}Centro Fermi - Museo Storico della Fisica e Centro Studi e Ricerche ``Enrico Fermi', Rome, Italy
\item \Idef{org11}Chicago State University, Chicago, Illinois, United States
\item \Idef{org12}China Institute of Atomic Energy, Beijing, China
\item \Idef{org13}Comenius University Bratislava, Faculty of Mathematics, Physics and Informatics, Bratislava, Slovakia
\item \Idef{org14}COMSATS University Islamabad, Islamabad, Pakistan
\item \Idef{org15}Creighton University, Omaha, Nebraska, United States
\item \Idef{org16}Department of Physics, Aligarh Muslim University, Aligarh, India
\item \Idef{org17}Department of Physics, Pusan National University, Pusan, Republic of Korea
\item \Idef{org18}Department of Physics, Sejong University, Seoul, Republic of Korea
\item \Idef{org19}Department of Physics, University of California, Berkeley, California, United States
\item \Idef{org20}Department of Physics, University of Oslo, Oslo, Norway
\item \Idef{org21}Department of Physics and Technology, University of Bergen, Bergen, Norway
\item \Idef{org22}Dipartimento di Fisica dell'Universit\`{a} 'La Sapienza' and Sezione INFN, Rome, Italy
\item \Idef{org23}Dipartimento di Fisica dell'Universit\`{a} and Sezione INFN, Cagliari, Italy
\item \Idef{org24}Dipartimento di Fisica dell'Universit\`{a} and Sezione INFN, Trieste, Italy
\item \Idef{org25}Dipartimento di Fisica dell'Universit\`{a} and Sezione INFN, Turin, Italy
\item \Idef{org26}Dipartimento di Fisica e Astronomia dell'Universit\`{a} and Sezione INFN, Bologna, Italy
\item \Idef{org27}Dipartimento di Fisica e Astronomia dell'Universit\`{a} and Sezione INFN, Catania, Italy
\item \Idef{org28}Dipartimento di Fisica e Astronomia dell'Universit\`{a} and Sezione INFN, Padova, Italy
\item \Idef{org29}Dipartimento di Fisica `E.R.~Caianiello' dell'Universit\`{a} and Gruppo Collegato INFN, Salerno, Italy
\item \Idef{org30}Dipartimento DISAT del Politecnico and Sezione INFN, Turin, Italy
\item \Idef{org31}Dipartimento di Scienze e Innovazione Tecnologica dell'Universit\`{a} del Piemonte Orientale and INFN Sezione di Torino, Alessandria, Italy
\item \Idef{org32}Dipartimento di Scienze MIFT, Universit\`{a} di Messina, Messina, Italy
\item \Idef{org33}Dipartimento Interateneo di Fisica `M.~Merlin' and Sezione INFN, Bari, Italy
\item \Idef{org34}European Organization for Nuclear Research (CERN), Geneva, Switzerland
\item \Idef{org35}Faculty of Electrical Engineering, Mechanical Engineering and Naval Architecture, University of Split, Split, Croatia
\item \Idef{org36}Faculty of Engineering and Science, Western Norway University of Applied Sciences, Bergen, Norway
\item \Idef{org37}Faculty of Nuclear Sciences and Physical Engineering, Czech Technical University in Prague, Prague, Czech Republic
\item \Idef{org38}Faculty of Science, P.J.~\v{S}af\'{a}rik University, Ko\v{s}ice, Slovakia
\item \Idef{org39}Frankfurt Institute for Advanced Studies, Johann Wolfgang Goethe-Universit\"{a}t Frankfurt, Frankfurt, Germany
\item \Idef{org40}Fudan University, Shanghai, China
\item \Idef{org41}Gangneung-Wonju National University, Gangneung, Republic of Korea
\item \Idef{org42}Gauhati University, Department of Physics, Guwahati, India
\item \Idef{org43}Helmholtz-Institut f\"{u}r Strahlen- und Kernphysik, Rheinische Friedrich-Wilhelms-Universit\"{a}t Bonn, Bonn, Germany
\item \Idef{org44}Helsinki Institute of Physics (HIP), Helsinki, Finland
\item \Idef{org45}High Energy Physics Group,  Universidad Aut\'{o}noma de Puebla, Puebla, Mexico
\item \Idef{org46}Hiroshima University, Hiroshima, Japan
\item \Idef{org47}Hochschule Worms, Zentrum  f\"{u}r Technologietransfer und Telekommunikation (ZTT), Worms, Germany
\item \Idef{org48}Horia Hulubei National Institute of Physics and Nuclear Engineering, Bucharest, Romania
\item \Idef{org49}Indian Institute of Technology Bombay (IIT), Mumbai, India
\item \Idef{org50}Indian Institute of Technology Indore, Indore, India
\item \Idef{org51}Indonesian Institute of Sciences, Jakarta, Indonesia
\item \Idef{org52}INFN, Laboratori Nazionali di Frascati, Frascati, Italy
\item \Idef{org53}INFN, Sezione di Bari, Bari, Italy
\item \Idef{org54}INFN, Sezione di Bologna, Bologna, Italy
\item \Idef{org55}INFN, Sezione di Cagliari, Cagliari, Italy
\item \Idef{org56}INFN, Sezione di Catania, Catania, Italy
\item \Idef{org57}INFN, Sezione di Padova, Padova, Italy
\item \Idef{org58}INFN, Sezione di Roma, Rome, Italy
\item \Idef{org59}INFN, Sezione di Torino, Turin, Italy
\item \Idef{org60}INFN, Sezione di Trieste, Trieste, Italy
\item \Idef{org61}Inha University, Incheon, Republic of Korea
\item \Idef{org62}Institute for Nuclear Research, Academy of Sciences, Moscow, Russia
\item \Idef{org63}Institute for Subatomic Physics, Utrecht University/Nikhef, Utrecht, Netherlands
\item \Idef{org64}Institute of Experimental Physics, Slovak Academy of Sciences, Ko\v{s}ice, Slovakia
\item \Idef{org65}Institute of Physics, Homi Bhabha National Institute, Bhubaneswar, India
\item \Idef{org66}Institute of Physics of the Czech Academy of Sciences, Prague, Czech Republic
\item \Idef{org67}Institute of Space Science (ISS), Bucharest, Romania
\item \Idef{org68}Institut f\"{u}r Kernphysik, Johann Wolfgang Goethe-Universit\"{a}t Frankfurt, Frankfurt, Germany
\item \Idef{org69}Instituto de Ciencias Nucleares, Universidad Nacional Aut\'{o}noma de M\'{e}xico, Mexico City, Mexico
\item \Idef{org70}Instituto de F\'{i}sica, Universidade Federal do Rio Grande do Sul (UFRGS), Porto Alegre, Brazil
\item \Idef{org71}Instituto de F\'{\i}sica, Universidad Nacional Aut\'{o}noma de M\'{e}xico, Mexico City, Mexico
\item \Idef{org72}iThemba LABS, National Research Foundation, Somerset West, South Africa
\item \Idef{org73}Jeonbuk National University, Jeonju, Republic of Korea
\item \Idef{org74}Johann-Wolfgang-Goethe Universit\"{a}t Frankfurt Institut f\"{u}r Informatik, Fachbereich Informatik und Mathematik, Frankfurt, Germany
\item \Idef{org75}Joint Institute for Nuclear Research (JINR), Dubna, Russia
\item \Idef{org76}Korea Institute of Science and Technology Information, Daejeon, Republic of Korea
\item \Idef{org77}KTO Karatay University, Konya, Turkey
\item \Idef{org78}Laboratoire de Physique des 2 Infinis, Ir\`{e}ne Joliot-Curie, Orsay, France
\item \Idef{org79}Laboratoire de Physique Subatomique et de Cosmologie, Universit\'{e} Grenoble-Alpes, CNRS-IN2P3, Grenoble, France
\item \Idef{org80}Lawrence Berkeley National Laboratory, Berkeley, California, United States
\item \Idef{org81}Lund University Department of Physics, Division of Particle Physics, Lund, Sweden
\item \Idef{org82}Nagasaki Institute of Applied Science, Nagasaki, Japan
\item \Idef{org83}Nara Women{'}s University (NWU), Nara, Japan
\item \Idef{org84}National and Kapodistrian University of Athens, School of Science, Department of Physics , Athens, Greece
\item \Idef{org85}National Centre for Nuclear Research, Warsaw, Poland
\item \Idef{org86}National Institute of Science Education and Research, Homi Bhabha National Institute, Jatni, India
\item \Idef{org87}National Nuclear Research Center, Baku, Azerbaijan
\item \Idef{org88}National Research Centre Kurchatov Institute, Moscow, Russia
\item \Idef{org89}Niels Bohr Institute, University of Copenhagen, Copenhagen, Denmark
\item \Idef{org90}Nikhef, National institute for subatomic physics, Amsterdam, Netherlands
\item \Idef{org91}NRC Kurchatov Institute IHEP, Protvino, Russia
\item \Idef{org92}NRC \guillemotleft Kurchatov\guillemotright~Institute - ITEP, Moscow, Russia
\item \Idef{org93}NRNU Moscow Engineering Physics Institute, Moscow, Russia
\item \Idef{org94}Nuclear Physics Group, STFC Daresbury Laboratory, Daresbury, United Kingdom
\item \Idef{org95}Nuclear Physics Institute of the Czech Academy of Sciences, \v{R}e\v{z} u Prahy, Czech Republic
\item \Idef{org96}Oak Ridge National Laboratory, Oak Ridge, Tennessee, United States
\item \Idef{org97}Ohio State University, Columbus, Ohio, United States
\item \Idef{org98}Petersburg Nuclear Physics Institute, Gatchina, Russia
\item \Idef{org99}Physics department, Faculty of science, University of Zagreb, Zagreb, Croatia
\item \Idef{org100}Physics Department, Panjab University, Chandigarh, India
\item \Idef{org101}Physics Department, University of Jammu, Jammu, India
\item \Idef{org102}Physics Department, University of Rajasthan, Jaipur, India
\item \Idef{org103}Physikalisches Institut, Eberhard-Karls-Universit\"{a}t T\"{u}bingen, T\"{u}bingen, Germany
\item \Idef{org104}Physikalisches Institut, Ruprecht-Karls-Universit\"{a}t Heidelberg, Heidelberg, Germany
\item \Idef{org105}Physik Department, Technische Universit\"{a}t M\"{u}nchen, Munich, Germany
\item \Idef{org106}Politecnico di Bari, Bari, Italy
\item \Idef{org107}Research Division and ExtreMe Matter Institute EMMI, GSI Helmholtzzentrum f\"ur Schwerionenforschung GmbH, Darmstadt, Germany
\item \Idef{org108}Rudjer Bo\v{s}kovi\'{c} Institute, Zagreb, Croatia
\item \Idef{org109}Russian Federal Nuclear Center (VNIIEF), Sarov, Russia
\item \Idef{org110}Saha Institute of Nuclear Physics, Homi Bhabha National Institute, Kolkata, India
\item \Idef{org111}School of Physics and Astronomy, University of Birmingham, Birmingham, United Kingdom
\item \Idef{org112}Secci\'{o}n F\'{\i}sica, Departamento de Ciencias, Pontificia Universidad Cat\'{o}lica del Per\'{u}, Lima, Peru
\item \Idef{org113}St. Petersburg State University, St. Petersburg, Russia
\item \Idef{org114}Stefan Meyer Institut f\"{u}r Subatomare Physik (SMI), Vienna, Austria
\item \Idef{org115}SUBATECH, IMT Atlantique, Universit\'{e} de Nantes, CNRS-IN2P3, Nantes, France
\item \Idef{org116}Suranaree University of Technology, Nakhon Ratchasima, Thailand
\item \Idef{org117}Technical University of Ko\v{s}ice, Ko\v{s}ice, Slovakia
\item \Idef{org118}The Henryk Niewodniczanski Institute of Nuclear Physics, Polish Academy of Sciences, Cracow, Poland
\item \Idef{org119}The University of Texas at Austin, Austin, Texas, United States
\item \Idef{org120}Universidad Aut\'{o}noma de Sinaloa, Culiac\'{a}n, Mexico
\item \Idef{org121}Universidade de S\~{a}o Paulo (USP), S\~{a}o Paulo, Brazil
\item \Idef{org122}Universidade Estadual de Campinas (UNICAMP), Campinas, Brazil
\item \Idef{org123}Universidade Federal do ABC, Santo Andre, Brazil
\item \Idef{org124}University of Cape Town, Cape Town, South Africa
\item \Idef{org125}University of Houston, Houston, Texas, United States
\item \Idef{org126}University of Jyv\"{a}skyl\"{a}, Jyv\"{a}skyl\"{a}, Finland
\item \Idef{org127}University of Liverpool, Liverpool, United Kingdom
\item \Idef{org128}University of Science and Technology of China, Hefei, China
\item \Idef{org129}University of South-Eastern Norway, Tonsberg, Norway
\item \Idef{org130}University of Tennessee, Knoxville, Tennessee, United States
\item \Idef{org131}University of the Witwatersrand, Johannesburg, South Africa
\item \Idef{org132}University of Tokyo, Tokyo, Japan
\item \Idef{org133}University of Tsukuba, Tsukuba, Japan
\item \Idef{org134}Universit\'{e} Clermont Auvergne, CNRS/IN2P3, LPC, Clermont-Ferrand, France
\item \Idef{org135}Universit\'{e} de Lyon, Universit\'{e} Lyon 1, CNRS/IN2P3, IPN-Lyon, Villeurbanne, Lyon, France
\item \Idef{org136}Universit\'{e} de Strasbourg, CNRS, IPHC UMR 7178, F-67000 Strasbourg, France, Strasbourg, France
\item \Idef{org137}Universit\'{e} Paris-Saclay Centre d'Etudes de Saclay (CEA), IRFU, D\'{e}partment de Physique Nucl\'{e}aire (DPhN), Saclay, France
\item \Idef{org138}Universit\`{a} degli Studi di Foggia, Foggia, Italy
\item \Idef{org139}Universit\`{a} degli Studi di Pavia, Pavia, Italy
\item \Idef{org140}Universit\`{a} di Brescia, Brescia, Italy
\item \Idef{org141}Variable Energy Cyclotron Centre, Homi Bhabha National Institute, Kolkata, India
\item \Idef{org142}Warsaw University of Technology, Warsaw, Poland
\item \Idef{org143}Wayne State University, Detroit, Michigan, United States
\item \Idef{org144}Westf\"{a}lische Wilhelms-Universit\"{a}t M\"{u}nster, Institut f\"{u}r Kernphysik, M\"{u}nster, Germany
\item \Idef{org145}Wigner Research Centre for Physics, Budapest, Hungary
\item \Idef{org146}Yale University, New Haven, Connecticut, United States
\item \Idef{org147}Yonsei University, Seoul, Republic of Korea
\end{Authlist}
\endgroup
  %%%%%%% done by webmaster team

%%%%%%%%% appendix with author list
\newpage
\appendix
%
%% Following lines needed so, for instance, Fig D.1(a) not printed as simply 1(a) when referenced
\renewcommand{\thesubfigure}{\thefigure(\alph{subfigure})}
\makeatletter
\renewcommand{\p@subfigure}{}
\renewcommand{\@thesubfigure}{(\alph{subfigure})\hskip\subfiglabelskip}
%\input{}               %%%%%%%%%%% put your appendices here
%

%************************************************************************************************************************
%************************************************************************************************************************
\section{Stavinskiy reference method}
\label{App:StavMethod}

Another option for obtaining the reference distribution, $B(k^{*})$, is to use, what will be referred to as, the ``Stavinskiy method"~\cite{Stavinskiy04}.
The method was first proposed to handle the case of one event femtoscopy, and has been suggested for use in eliminating momentum conservation effects in the reference distribution~\cite{Lisa:2005dd}.
The method is appropriate for collisions between symmetric projectiles, at sufficiently large energy, with a detector which is symmetrical with respect to the transformation $\mathbf{r} \rightarrow \mathbf{-r}$.
The use of this method in a three-dimensional analysis of two-pion correlations produced, in comparison to the event mixing results, an increase of 6\% for $R_{\mathrm{side}}$ at low-$k_{\mathrm{T}}$ and up to 4\% for $R_{\mathrm{out}}$ and $R_{\mathrm{long}}$~\cite{Aamodt:2011mr}.
The purpose of using the Stavinskiy method in this \LamK analysis is to rid the correlation functions of the non-femtoscopic background.  
More specifically, the intent is to handle background contributions from elliptic flow, and other sources having reflection symmetry in the transverse plane.  
With the Stavinskiy method, mixed-event pairs are not used for the reference distribution; instead, same-event pseudo-pairs, formed by rotating one particle in a real pair by 180$^\circ$ in the transverse plane, are used~\cite{PhysRevD.82.052001}.  
This rotation rids the pairs of any femtoscopic correlation, while maintaining correlations due to elliptic flow (and other suitably symmetric contributors).
Care needs to be taken in treating the pseudo-pairs exactly like the real pairs; e.g., the pseudo-pairs should be exposed to the same pair rejection procedures used in the analysis on the real pairs.
The results of correctly implementing such a procedure are shown in Fig.~\ref{fig:StavCfs_Correct_LamKchP}.
The figure demonstrates, for the \LamKchP system, that the Stavinskiy method is effective in flattening the correlation function in the region where no femtoscopic signal is expected.
This procedure flattens the non-femtoscopic background equally well for the \LamKchM system, but is less effective for the \LamKs system.

\begin{figure}[h!]
  \centering
  \includegraphics[width=0.50\textwidth]{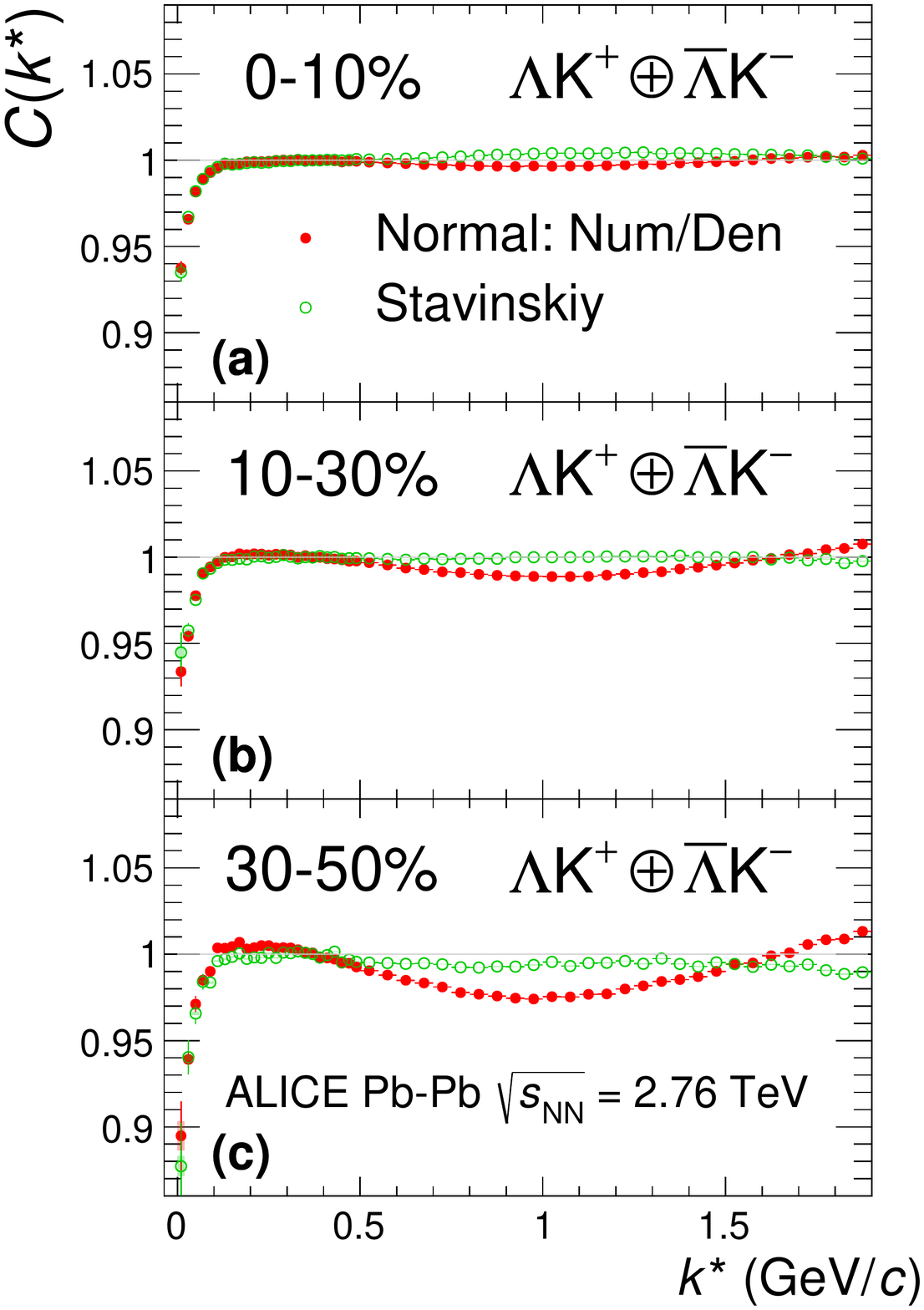}
  \caption[\LamKchP Stavinskiy Correlation Functions]
  {
  (Color online) Correlation functions for the $\Lambda\mathrm{K^{+}}\oplus\overline{\Lambda}\mathrm{K^{-}}$ system built using the Stavinskiy method for 0--10\%, 10--30\%, and 30--50\% centrality intervals.  Closed symbols represent correlations built using the normal mixed-event reference distribution, while open symbols represent correlations formed using the Stavinskiy same-event pseudo-pairs as a reference.
  Statistical (lines) and systematic (boxes) uncertainties are shown.
  }
  \label{fig:StavCfs_Correct_LamKchP}
\end{figure}

%************************************************************************************************************************
%************************************************************************************************************************
\section{Strong and Coulomb fitter}
\label{App:CoulombFitter}

When modeling systems which include both strong and Coulomb effects, Eq.~(\ref{eqn:LednickyEqn}) is no longer valid, and there exists no analytical form with which to fit.
To model such a system, a more fundamental approach must be taken, beginning with Eq.~(\ref{eqn:KooninPrattEqn}) and using the two-particle wave-function which includes both strong and Coulomb interactions~\cite{Lednicky:2005tb},
\begin{equation}
 \Psi_{\mathbf{k^{*}}}(\mathbf{r^{*}}) = e^{i\delta_{\mathrm{c}}}\sqrt{A_{\mathrm{c}}(\eta)}[e^{i\mathbf{k^{*}} \times \mathbf{r^{*}}}F(-i\eta,1,i\xi) + f_{\mathrm{c}}(k^{*})\frac{\tilde{G}(\rho,\eta)}{r^{*}}],
\label{eqn:CoulombWaveFcn}
\end{equation}
where $\rho = k^{*}r^{*}$, $\eta = (k^{*}a_{\mathrm{c}})^{-1}$, $\xi = \mathbf{k^{*}} \times \mathbf{r^{*}} + k^{*}r^{*} \equiv \rho(1+\cos\theta^{*})$, and $a_{\mathrm{c}} = (\mu z_{1}z_{2}e^{2})^{-1}$ is the two-particle Bohr radius (including the sign of the interaction).  
Furthermore, $\delta_{\mathrm{c}}$ is the Coulomb s-wave phase shift, $A_{\mathrm{c}}(\eta)$ is the Coulomb penetration factor, $\tilde{G} = \sqrt{A_{c}}(G_{0} + iF_{0})$ is a combination of the regular ($F_{0}$) and singular ($G_{0}$) s-wave Coulomb functions.  
Finally, $f_{\mathrm{c}}(k^{*})$ is the s-wave scattering amplitude,
\begin{equation}
 f_{\mathrm{c}}(k^{*}) = \left[\frac{1}{f_{0}} + \frac{1}{2}d_{0}k^{*2} - \frac{2}{a_{\mathrm{c}}}h(\eta) - ik^{*}A_{\mathrm{c}}(\eta)\right]^{-1},
\label{eqn:CoulombScattAmp}
\end{equation}
where the ``h-function", $h(\eta$), is expressed through the digamma function, $\psi(z)$ = $\Gamma'(z)/\Gamma(z)$ as
\begin{equation}
 h(\eta) = 0.5[\psi(i\eta) + \psi(-i\eta) - \ln(\eta^{2})].
\label{eqn:LednickyHFunction}
\end{equation} 
In this case, the $\lambda$ parameter may be included as
\begin{equation}
 C(\mathbf{k^{*}}) = (1 - \lambda) + \lambda\int S(\mathbf{r^{*}})|\Psi^{S}_{\mathbf{k^{*}}}(\mathbf{r^{*}})|^{2}\mathrm{d}^{3}r^{*}.
\label{eqn:GenCfEqnwLambda}
\end{equation}
To build a fit function for a system including both strong and Coulomb interactions two related options were considered. 
The first option was to numerically integrate Eq.~(\ref{eqn:KooninPrattEqn}).  
The second option was to simulate a large sample of particle pairs, calculate the wave function describing the interaction, and average to obtain the integral in Eq.~(\ref{eqn:KooninPrattEqn}). 
For this analysis, the latter option was adopted.

%************************************************************************************************************************
%************************************************************************************************************************
\section{Relative emission shifts with THERMINATOR 2}
\label{App:THERM}

Figure~\ref{fig:LamKchP_StdThermSources} shows \LamKchP results from the THERMINATOR 2 event generator for an impact parameter of $b = 2$ fm.
As THERMINATOR 2 does not include any final state effects, the femtoscopic correlation was introduced by assuming a set of scattering parameters ($\Re f_{0},\, \Im f_{0},\, d_{0}$) = ($-$0.60 fm, 0.51 fm, 0.83 fm) and weighting the pairs in the signal distribution with the modulus squared of the two-particle wave function, $|\Psi|^{2}$.

The top row of Fig.~\ref{fig:LamKchP_StdThermSources} shows the experimental $\Lambda\mathrm{K}^{+}\oplus\overline{\Lambda}\mathrm{K}^{-}$ data together with the simulation results (a) for the one-dimensional correlation function and (b) for the real part of the $l=1$, $m=1$ component, $\Re C_{11}$, of the spherical harmonic decomposition.
The other four plots in Fig.~\ref{fig:LamKchP_StdThermSources} show the two-particle emission function (i.e., the pair separation distributions) from the simulation in the (c) out ($r^{*}_{\mathrm{out}}$), (d) side ($r^{*}_{\mathrm{side}}$), and (e) long ($r^{*}_{\mathrm{long}}$) directions, as well as (f) the temporal characteristics of the source ($\Delta t^{*}$), all measured in the PRF.
The source distributions have all been fitted with a Gaussian form, the results of which are printed within the respective plots.
One immediately sees a significant spatial shift in the out direction, $\mu_{\mathrm{out}} \approx$ 4 fm, and negligible shift in the other two directions, $\mu_{\mathrm{side}} \approx \mu_{\mathrm{long}} \approx$ 0 fm.
In other words, the figure demonstrates that, within the THERMINATOR 2 model, the \Lam is, on average, emitted further out than its K partner.
Additionally, the figure shows a nonzero temporal shift, $\mu_{\Delta t} \approx$ $-$2.7 fm/$c$, signifying that the \Lam is, on average, emitted earlier than its K partner within the model.

\begin{figure}[h!]
  \centering
  \includegraphics[width=0.8\textwidth]{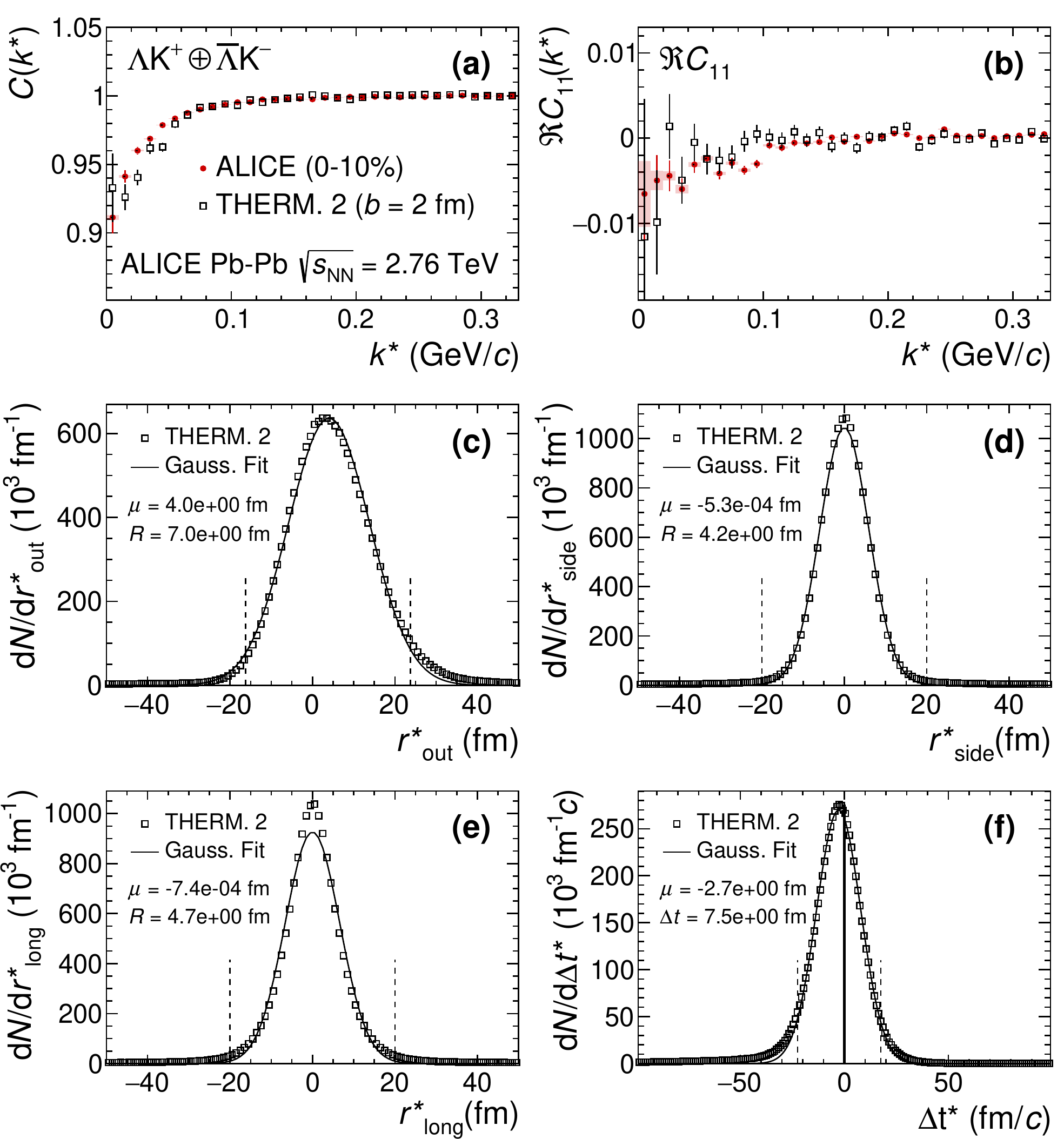}
  \caption[THERMINATOR 2 simulation for \LamKchP]
  {
  Results from the THERMINATOR 2 simulation implemented with an impact parameter $b = 2$ fm for the \LamKchP pair system.
  Where experimental data are shown, lines represent statistical uncertainties, while boxes represent systematic uncertainties.
  (a) the one-dimensional correlation function from THERMINATOR 2 together with the experimental data.
  (b) the $\Re C_{11}$ component of a spherical harmonic decomposition of the THERMINATOR 2 simulation together with the experimental data.
  The other four panels show the source distribution from the simulation in the (c) out, (d) side, and (e) long directions, as well as (f) the temporal characteristics, all in the PRF.
  The source distributions have all been fitted with a Gaussian form over the regions contained within the dashed lines, the results of which are printed within the respective plots.
  }
  \label{fig:LamKchP_StdThermSources}
\end{figure}

This section concludes with a brief look at how a spatial separation of the single particle sources affects the radii extracted from a femtoscopic analysis.
To achieve this, THERMINATOR 2 is used in a similar fashion as described above, but with one important difference.
Instead of taking the source information from THERMINATOR 2, the source is drawn from a pre-determined Gaussian distribution.
In all, $R_{\mathrm{out}} = R_{\mathrm{side}} = R_{\mathrm{long}}$ = 5 fm, and $\mu_{\mathrm{side}} = \mu_{\mathrm{long}}$ = 0 fm.
The cases of $\mu_{\mathrm{out}}$ = 0 fm, $\mu_{\mathrm{out}}$ = 1 fm, $\mu_{\mathrm{out}}$ = 3 fm, and $\mu_{\mathrm{out}}$ = 6 fm were studied within the simulation.
Note, within this implementation there is no time difference in the emission of the \Lam and K particles.
For each, a one-dimensional correlation function is generated and fit with the Lednick\'y model, as shown in Fig.~\ref{fig:LamKchP_ThermSources_VaryMuOut}.
The scattering parameters are known precisely here, as they served as the weights used in the simulation, and are kept constant in the fit.
The real component of the scattering length describes the effect of the strong interaction, and is very narrow in \kstar (\kstar $\lesssim$ 100 MeV/$c$).
The imaginary component of the scattering length accounts for inelastic scattering channels, and produces a wide (hundreds of MeV/$c$) negative correlation which goes to zero at \kstar= 0 MeV/$c$.
The interplay of these two components determines the final shape of the correlation function; in the case of Fig.~\ref{fig:LamKchP_ThermSources_VaryMuOut}, although both $\Re(f_{0})$ and $\Im(f_{0})$ lead to a suppression, the combination of these distinct contributions results in a rise of the correlation function at low \kstar (while still remaining less than unity).
For the fit, only the extracted one-dimensional source size is of interest here, so the $\lambda$ parameter is also fixed at unity.
The figure demonstrates that as the separation $\mu_{\mathrm{out}}$ increases, so do the extracted femtoscopic radii.
Figure~\ref{fig:LamKchP_ThermSources_VaryMuOut_SH} shows, together with the experimental \LamKchP data, the effect of increasing $\mu_{\mathrm{out}}$ on the spherical harmonic $l=0$, $m=0$ component, $C_{00}$, and on the real part of the $l=1$, $m=1$ component, $\Re C_{11}$.
The figures shows that as $\mu_{\mathrm{out}}$ increases, so does the magnitude of the $\Re C_{11}$ signal.

\begin{figure}[h]
  \centering
  \includegraphics[width=\textwidth]{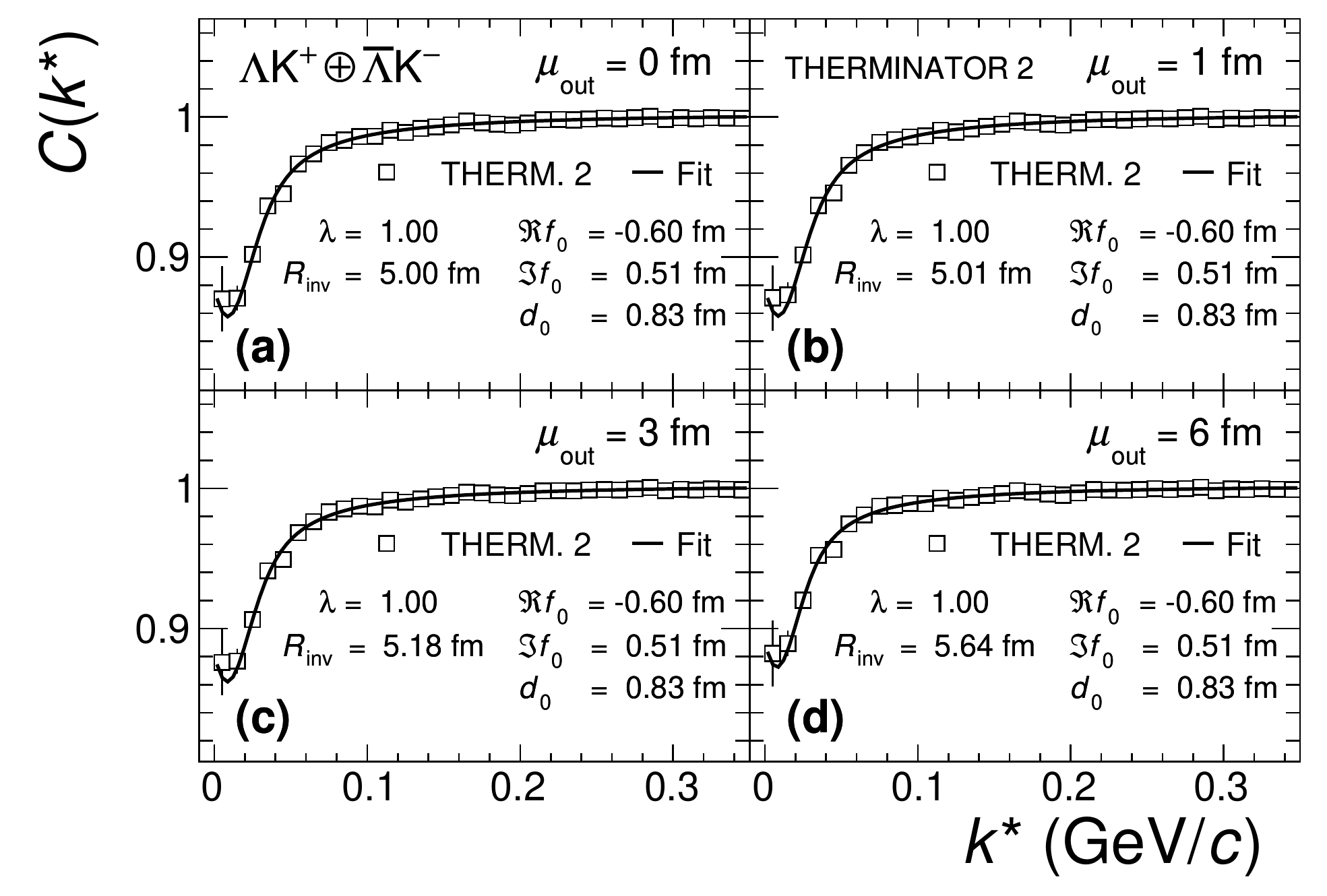}
  \caption[Varying $\mu_{\mathrm{Out}}$ with THERMINATOR 2]
  {
  Probing the effect of varying the source shift in the outward direction, $\mu_{\mathrm{out}}$, within the THERMINATOR 2 framework.  
  To achieve this, particle pairs are formed from the simulation, but with altered spatial characteristics achieved by drawing the out, side, and long components from predetermined Gaussian distributions.  
  The sources in all three directions are Gaussians of width 5 fm.
  The distributions used for the side and long direction are centered at the origin, while the shift in the outward direction, $\mu_{\mathrm{out}}$, is varied.
  The plots show fits resulting from sources with $\mu_{\mathrm{out}}$ increasing from 0 to 6 fm. 
  The effect of increasing $\mu_{\mathrm{out}}$ clearly increases the effective radius extracted in the fit.
  }
  \label{fig:LamKchP_ThermSources_VaryMuOut}
\end{figure}

\begin{figure}[htp]
  \centering
  %%----start of first subfigure---
  \subfigure{
    \label{fig:LamKchP_ThermSources_VaryMuOut_SH:a}
    \includegraphics[width=0.49\linewidth]{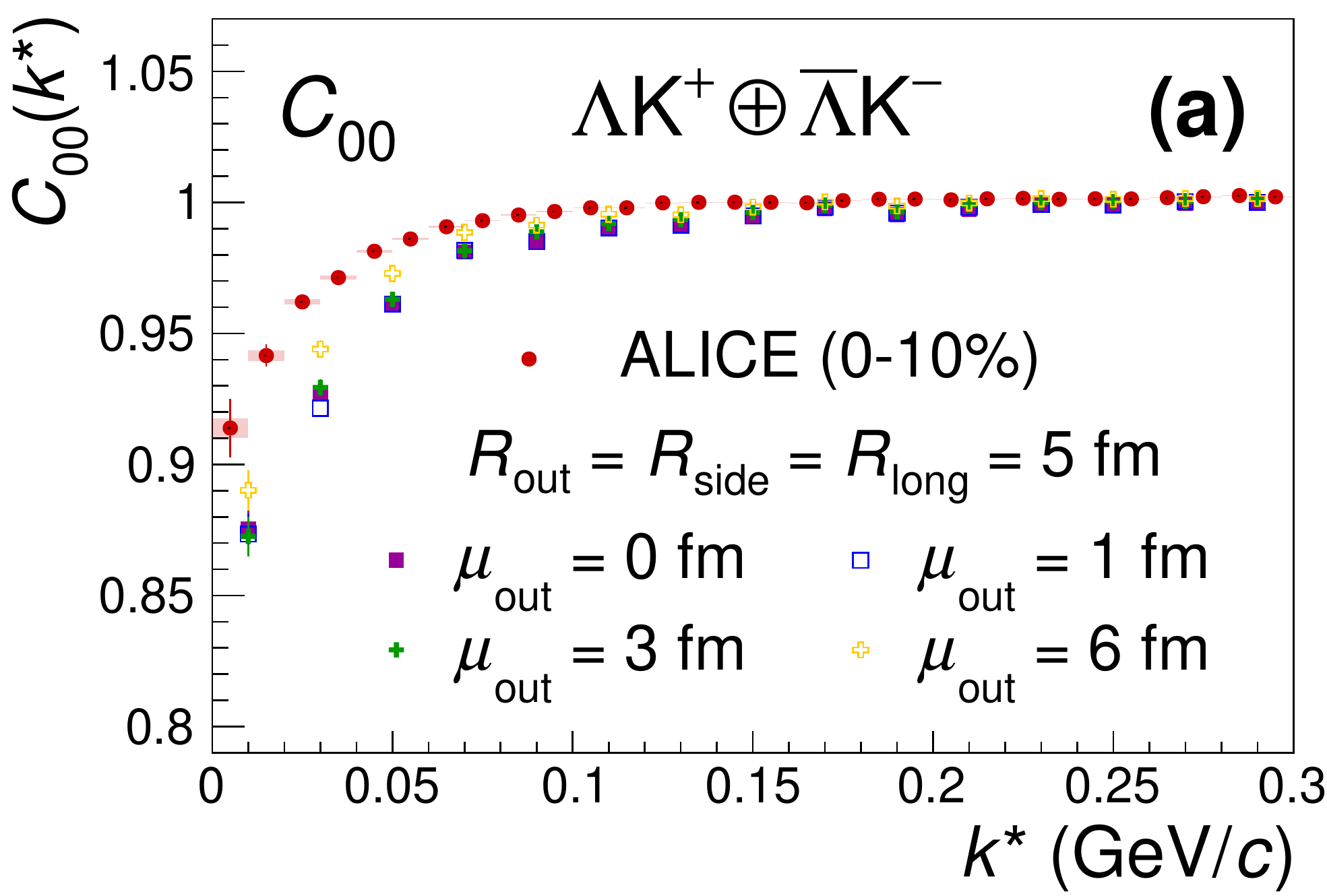}}
  %%----start of second subfigure---  
  \subfigure{
    \label{fig:LamKchP_ThermSources_VaryMuOut_SH:b}
    \includegraphics[width=0.49\linewidth]{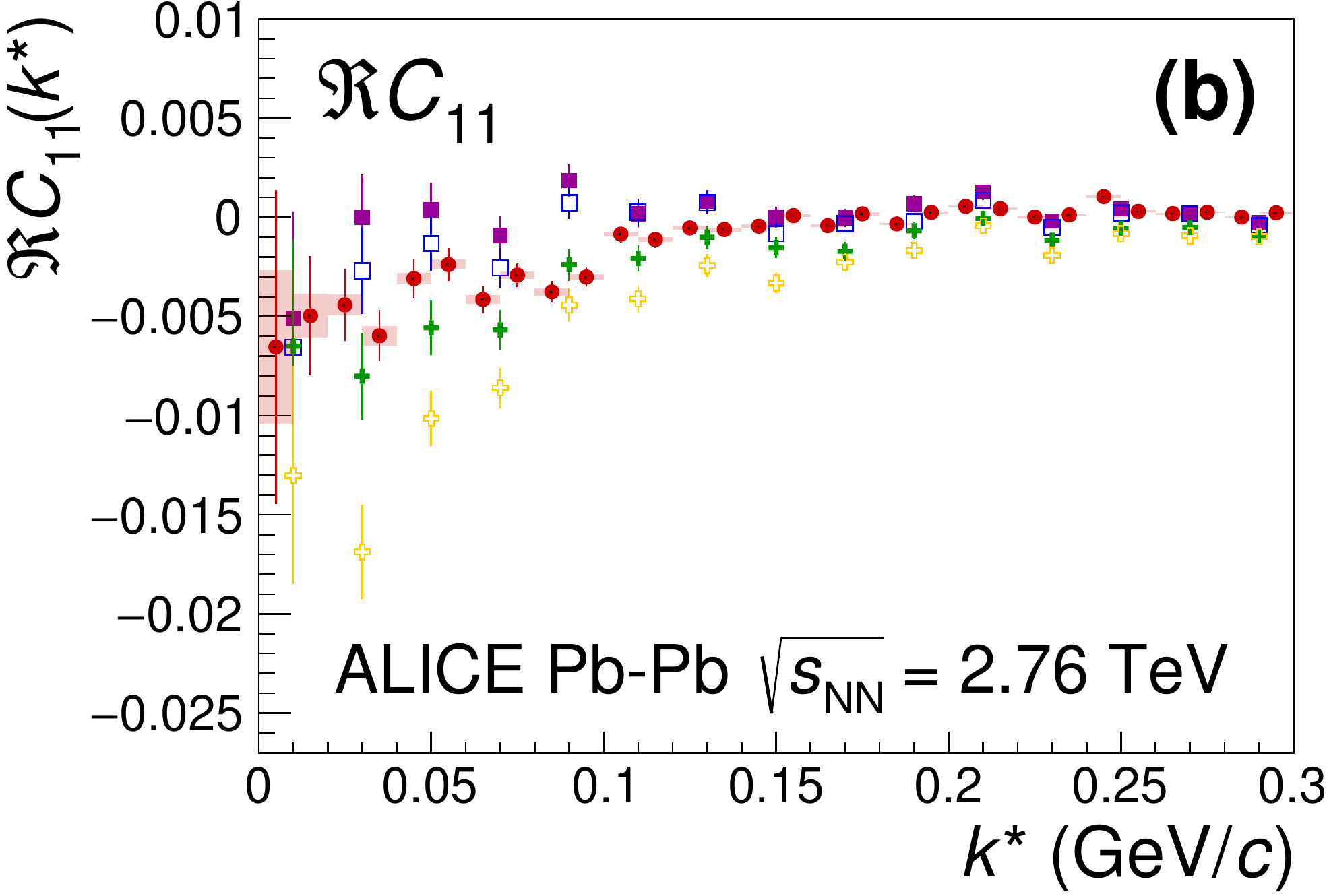}}
  %%----overall caption----
  \caption[\LamKchP $C_{00}$ and $\Re C_{11}$ Spherical Harmonic Components (0--10\%) with THERMINATOR 2 ($b = 2$ fm]
  {
  (Color online) Spherical harmonics components (left) $C_{00}$ and (right) $\Re C_{11}$ of the \LamKchP correlation function for the 0--10\% centrality interval shown with results from the THERMINATOR 2 simulation implemented with different shifts in the outward direction, $\mu_{\mathrm{out}}$, as described in the text.
  Statistical (lines) and systematic (boxes) uncertainties are shown for the experimental data.
  }  
  \label{fig:LamKchP_ThermSources_VaryMuOut_SH}
\end{figure}

\end{document}